\documentclass{article}

\usepackage{amsfonts}
\usepackage{amssymb,epic,eepic}
\usepackage{amsmath}
\usepackage{dcolumn}
\usepackage{bm}
\newcommand{\hr}{{\mathcal H}}
\newcommand{\cn}{{\mathcal N }}
\newcommand{\E}{{\mathcal E }}

\newcommand{\cs}{{\mathcal S}}
\newcommand{\crr}{{\mathcal R}}
\newcommand{\fr}{{\mathcal F}}
\newcommand{\gr}{{\mathcal G}}
\newcommand{\fri}{{\mathfrak I}}
\newcommand{\kr}{{\mathcal K}}

\newcommand{\cc}{{\mathbb C}}
\newcommand{\rr}{{\mathbb R}}
\newcommand{\nn}{{\mathbb N}}
\newcommand{\idn}{\mathbf{1}}

\newcommand{\eps}{{\varepsilon}}        
\newcommand{\vphi}{{\varphi}}           

\newcommand{\A}{\mathcal A}

\newcommand{\D}{\mathcal D}

\newtheorem{theorem}{Theorem}

\newtheorem{lemma}[theorem]{Lemma}

\newtheorem{remark}[theorem]{Remark}

\newcommand{\U}{\mathcal U}
\newcommand{\V}{\mathcal V}
\newcommand{\tr}{\mathrm{tr}}
\newcommand{\supp}{\mathrm{supp}}

\begin{document}
\title{Entanglement transmission and generation under channel uncertainty: Universal quantum channel coding}
\author{Igor Bjelakovi\'c $^{1,2}$, Holger Boche $^{1,2}$, Janis N\"otzel $^{1}$\\
$^{1}$ {\footnotesize Heinrich-Hertz-Lehrstuhl f\"ur Mobilkommunikation (HFT 6)}\\
{\footnotesize Technische Universit\"at Berlin, Germany}\\
$^{2}${\footnotesize Institut f\"ur Mathematik, Technische Universit\"at Berlin, Germany }
}

\maketitle
\begin{abstract}
We determine the optimal rates of universal quantum codes for entanglement transmission and generation under channel uncertainty. In the simplest scenario the sender and receiver are provided merely with the information that the channel they use belongs to a given set of channels, so that they are forced to use quantum codes that are reliable for the whole set of channels. This is precisely the quantum analog of the compound channel coding problem. We determine the entanglement transmission and entanglement-generating capacities of compound quantum channels and show that they are equal. Moreover, we investigate two variants of that basic scenario, namely the cases of informed decoder or informed encoder, and derive corresponding capacity results.
\end{abstract}

\tableofcontents
\section{\label{sec:Introduction}Introduction}
The determination of capacities of quantum channels in various settings has been a field of intense work over the last decade. In contrast to classical information theory, to any quantum channel we can associate in a natural way different notions of capacity depending on what is to be transmitted over the channel and which figure of merit is chosen as the criterion for the success of the particular quantum communication task. For example we may try to determine the maximum number of classical messages that can be reliably distinguished at the output of the channel leading to the notion of classical capacity of a quantum channel. We might as well wish to establish secure classical communication over a quantum channel, giving rise to the definition of a channel's private capacity.\\
On the other hand, in the realm of quantum communication, one may ask e.g. the question what the maximal amount of entanglement is that we can generate or transmit over a given quantum channel, leading to the notions of entanglement-generating and entanglement transmission capacities. Other examples of quantum capacities are the subspace transmission and average subspace transmission capacities. Such quantum communication tasks are needed, for example, to support computation in quantum circuits or to provide the best possible supply of pure entanglement in a noisy environment. Fortunately, these genuinely quantum mechanical capacities are shown to be equal for perfectly known single user channels \cite{bkn}, \cite{kretschmann-werner}. \\
First results indicating that coherent information was to play a role in the determination of the quantum capacity of memoryless channels were established by Schumacher and Nielsen \cite{schumacher-nielsen} and, independently, by Lloyd \cite{lloyd} who was the first to conjecture that indeed the regularized coherent information would give the correct formula for the quantum capacity and gave strong heuristic evidence to his claim. In 1998 Barnum, Knill, and Nielsen and Barnum, Nielsen, and Schumacher \cite{bkn} gave the first upper bound on the capacity of a memoryless channel in terms of the regularized coherent information. Later on, Shor \cite{shor} and Devetak \cite{devetak} offered two independent approaches to the achievability part of the coding theorem. Despite the fact that the regularized coherent information was identified as the capacity of memoryless quantum channels many other approaches to the coding theorem have been offered subsequently, for example Devetak and Winter \cite{devetak-winter} and Hayden, Shor, and Winter \cite{hayden-shor-winter}. Of particular interest for our paper are the developments by Klesse \cite{klesse} and Hayden, Horodecki, Winter, and Yard \cite{hayden} based on the decoupling idea which can be traced back to Schumacher and Westmoreland \cite{schu-we-approx}. In fact, the main purpose of our work is to show that the decoupling idea can be utilized to prove the existence of reliable universal quantum codes for entanglement transmission and generation. \\
On the other hand, the classical capacity of memoryless quantum channels has been determined in the pioneering work by Holevo \cite{holevo} and Schumacher and Westmoreland \cite{schumacher-westmoreland}. Their results have been substantially sharpened by Winter \cite{winter} and Ogawa and Nagaoka \cite{ogawa-nagaoka} who gave independent proofs of the strong converse to the coding theorem.\\
However, most of the work done so far on quantum channel capacities relies on the assumption that the channel is \emph{perfectly known} to the sender and receiver. Such a requirement is hardly fulfilled in many situations. In this paper we consider compound quantum channels which are among the simplest non-trivial models with channel uncertainty. A rough description of this communication scenario is that the sender and receiver do not know the memoryless channel they have to use. The prior knowledge they have access to is merely that the actual channel belongs to a set $\fri$ of channels which in turn is known to the sender and receiver. It is important to notice that we impose no restrictions on the set $\fri$, i.e. it can be finite, countably-infinite or uncountable. Our intention is to identify the best rates of quantum codes for entanglement transmission and generation that are reliable for the whole set of channels $\fri$ simultaneously. This is, in some sense, a quantum channel counterpart of the universal quantum data compression result discovered by Jozsa and the Horodecki family \cite{jozsa-horodecki}.\\
While the classical capacity of compound quantum channels has been determined only recently in \cite{bb-compound}, in this paper we will focus on entanglement-generating and entanglement transmission capacities of compound quantum channels. Specifically we will determine both of them and show that they are equal. The investigation of their relation to other possible definitions of quantum capacity of compound quantum channels in spirit of \cite{bkn}, \cite{kretschmann-werner} will be given elsewhere.
\subsection{\label{subsec:Related Work}Related Work}
The capacity of compound channels in the classical setting was determined by Wolfowitz \cite{wolfowitz-compound,wolfowitz-book} and Blackwell, Breiman, and Thomasian \cite{blackwell}. The full coding theorem for transmission of classical information via compound quantum channels was proven in \cite{bb-compound}. Subsequently, Hayashi \cite{hayashi-universal} obtained a closely related result with a completely different proof technique based on the Schur-Weyl duality from representation theory and the packing lemma from \cite{csiszar}.\\
In our previous paper \cite{bbn-1} we determined the entanglement transmission capacity of \emph{finite} quantum compound channels (i.e. $|\fri |<\infty$). Moreover, we were able to prove the coding theorem for arbitrary $\fri$ with \emph{informed decoder}. It is important to remark here that we used a different notion of codes in \cite{bbn-1}, following \cite{klesse}, which is motivated by the theory of quantum error correction. In the cases of an informed decoder and uninformed users this change does not appear to be of importance. In the case of an informed encoder it is of crucial importance in the proof of the direct part of the coding result.\\
In our former paper, the strategy of proof was as follows. First, we derived a modification of Klesse's one-shot coding result \cite{klesse} that was adapted to arithmetic averages of channels. Application of this theorem combined with a discretization technique based on $\tau$-nets yielded the coding result for quantum compound channels with informed decoder and arbitrary $\fri$.\\
With the help of the channel-estimation technique developed by Datta and Dorlas \cite{datta-dorlas} we were able to show that in the case of a \emph{finite} compound channel it is asymptotically of no relevance if one spends the first $\lfloor\sqrt{l}\rfloor$ transmissions for channel estimation, thus turning an uninformed decoder into an informed decoder. Since for an informed decoder we had already proven the existence of good codes, we were able to obtain the full coding result in the case $|\fri |<\infty$.\\
Unfortunately, the speed at which one can gain channel knowledge using the channel estimation technique we employed is highly dependent on the number of channels. Due to this fact, the combination of channel estimation and approximation of general compound channels through finite ones did not seem to work in the other two cases.\\
In this paper, we use a more direct strategy. First, we derive one-shot coding results for \emph{finite} compound channels with uninformed users and informed encoder. In order to evaluate the dependence of the derived bounds on the block length we have to project onto typical subspaces of suitable output states of the individual channels. Therefore, it turns out that we effectively end up in the scenario with \emph{informed decoder}. Now, instead of employing a channel estimation strategy we study the impact of these projections onto the typical subspaces on the entanglement fidelity of the entire encoding-decoding procedure. It turns out that these projections can simply be removed without decreasing the entanglement fidelity too much and we have got a universal (i.e. uninformed) decoder for our coding problem. Then, again, using the discretization technique based on $\tau$-nets we can convert these results for finite $\fri$ to arbitrary compound quantum channels.\\
Another difference to our previous paper \cite{bbn-1} is that we determine the optimal rates in all the scenarios described above for entanglement generation over compound quantum channels and show that they coincide with the entanglement transmission capacities.
\subsection{\label{subsec:Outline}Outline}
Section \ref{sec:codes-and-capacity} contains the fundamental definitions of codes and capacities for entanglement transmission in all three different settings. Moreover, the reader can find there the statement of our main result.\\
It is followed by a section on one-shot results containing the one-shot result of Klesse \cite{klesse}, as well as our modifications thereof. The modified coding results guarantee the existence of unitary encodings as well as recovery operations for finite arithmetic averaged channels in all three different cases and establish a relation between the rate of the code and its entanglement fidelity. We also give an estimate relating the entanglement fidelity of a coding-decoding procedure to that of a disturbed version, where disturbance means that the application of the channel is followed by a projection.\\
With these one-shot results at hand, in Section \ref{sec:Direct Part of The Coding Theorem for Finitely many Channels} we are able to prove the existence of codes for entanglement transmission of sufficiently high rates and entanglement fidelity asymptotically approaching one exponentially fast in the case of finite compound channels.\\
Section \ref{sec:discretization} states the basic properties of finite size nets in the set of quantum channels. They are used to approximate general sets of quantum channels and provide the link between finite and general compound channels. The construction is such that their size depends polynomially on the approximation parameter.\\
We use the coding results for finite compound channels and the properties of finite nets in section \ref{sec:Direct Parts of the Coding Theorems for General Quantum Compound Channels} to derive sharp lower bounds on the entanglement transmission capacity of general compound channels. This section also contains variants of the BSST Lemma \cite{bsst} where BSST stands for Bennett, Shor, Smolin, and Thapliyal. The proofs rely heavily on the difference in the polynomial growth of nets versus exponentially fast convergence to entanglement fidelity one for the codes in the finite setting.\\
The next section \ref{sec:Converse Parts of the Coding Theorems for General Quantum Compound Channels} contains the converse parts of the coding theorems for general compound channels. Since the converse must hold for arbitrary encoding schemes and since we explicitly allow the code space to be larger than the input space of the channels, we deviate from the usual structure and instead employ the converse part for the case of entanglement generation that was developed by Devetak \cite{devetak}. We also use a recent continuity result due to Leung and Smith \cite{leung-smith} that connects the difference in coherent information between nearby channels.\\
In section \ref{sec:Continuity of Compound Capacity} we show, once again using the work of Leung and Smith \cite{leung-smith}, that the entanglement transmission capacities of compound quantum channels are continuous with respect to the Hausdorff metric.\\
In the final section \ref{sec:ent-generating} we apply the results obtained so far to determine the entanglement-generating capacities of compound quantum channels. It is not very surprising that it turns out that they coincide with their counterparts for entanglement transmission.

\subsection{\label{Notation and Conventions}Notation and Conventions}
All Hilbert spaces are assumed to have finite dimension and are over the field $\cc$. $\mathcal{S}(\hr)$ is the set of states, i.e. positive semi-definite operators with trace $1$ acting on the Hilbert space $\hr$. Pure states are given by projections onto one-dimensional subspaces. A vector of unit length spanning such a subspace will therefore be referred to as a state vector. To each subspace $\fr$ of $\hr$ we can associate unique projection $q_{\fr}$ whose range is the subspace $\fr$ and we write $\pi_{\fr}$ for the maximally mixed state on $\fr$, i.e. $\pi_{\fr}:=\frac{q_{\fr}}{\textrm{tr}(q_{\fr})}$.\\
The set of completely positive trace preserving (CPTP) maps
between the operator spaces $\mathcal{B}(\hr)$ and
$\mathcal{B}(\kr)$ is denoted by $\mathcal{C}(\hr,\kr)$. Thus $\hr$ plays the role of the input Hilbert space to the channel (traditionally owned by Alice) and $\kr$ is channel's output Hilbert space (usually in Bob's possession).
$\mathcal{C}^{\downarrow}(\hr,\kr)$ stands for the set of
completely positive trace decreasing maps between
$\mathcal{B}(\hr)$ and $\mathcal{B}(\kr)$. $\mathfrak{U}(\hr)$ will denote in what follows the group of unitary operators acting on $\hr$. For a Hilbert space $\gr\subset \hr$ we will always identify $\mathfrak{U}(\gr)$ with a subgroup of $\mathfrak{U}(\hr)$ in the canonical way. For any projection $q\in\mathcal{B}(\hr)$ we set $q^\perp:=\idn_{\hr}-q$. Each projection $q\in\mathcal{B}(\hr)$ defines a completely positive trace decreasing map $\mathcal{Q}$ given by $\mathcal{Q}(a):=qaq$ for all $a\in\mathcal{B}(\hr)$. In a similar fashion any $u\in \mathfrak{U}(\hr)$ defines a $\U\in \mathcal{C}(\hr,\hr)$ by $\U(a):=uau^{\ast}$ for $a\in\mathcal{B}(\hr)$.\\
We use the base two
logarithm which is denoted by $\log$. The von Neumann entropy of
a state $\rho\in\mathcal{S}(\hr)$ is given by
\[S(\rho):=-\textrm{tr}(\rho \log\rho).  \]
The coherent information for $\cn\in \mathcal{C}(\hr,\kr) $ and
$\rho\in\mathcal{S}(\hr)$ is defined by
\[I_c(\rho, \cn):=S(\cn (\rho))- S( (id_{\hr}\otimes \cn)(|\psi\rangle\langle \psi|)  ),  \]
where $\psi\in\hr\otimes \hr$ is an arbitrary purification of the state $\rho$. Following the usual conventions we let $S_e(\rho,\cn):=S( (id_{\hr}\otimes \cn)(|\psi\rangle\langle \psi|)  )$ denote the entropy exchange.
A useful equivalent definition of $I_c(\rho,\cn)$ is given in terms of $\cn\in\mathcal{C}(\hr,\kr)$ and the complementary channel $\cn'\in\mathcal{C}(\hr,\hr_e)$ where $\hr_e$ denotes the Hilbert space of the environment: Due to Stinespring's dilation theorem $\cn$ can be represented as $\cn(\rho)=\textrm{tr}_{\hr_e}(v\rho v^{\ast})$ for $\rho \in\mathcal{S}(\hr) $ where $v:\hr \to\kr \otimes \hr_e$ is a linear isometry. The complementary channel $\cn'\in\mathcal{C}(\hr,\hr_e)$ to $\cn$ is given by
\[\cn'(\rho):= \textrm{tr}_{\hr}(v\rho v^{\ast})\qquad (\rho \in\mathcal{S}(\hr)).  \]
The coherent information can then be written as
\[I_c(\rho, \cn)=S(\cn (\rho))- S(\cn'(\rho)). \]
As a measure of closeness between two states $\rho,\sigma\in\mathcal S(\hr)$ we use the fidelity $F(\rho,\sigma):=||\sqrt{\rho}\sqrt{\sigma}||^2_1$. The fidelity is symmetric in the input and for a pure state $\rho=|\phi\rangle\langle\phi|$ we have $F(|\phi\rangle\langle\phi|,\sigma)=\langle\phi,\sigma\phi\rangle$.\\
A closely related quantity is the entanglement fidelity. For $\rho\in\mathcal{S}(\hr)$ and $\cn\in
\mathcal{C}^{\downarrow}(\hr,\kr)$ it is given by
\[F_e(\rho,\cn):=\langle\psi, (id_{\hr}\otimes \cn)(|\psi\rangle\langle \psi|)     \psi\rangle,  \]
with $\psi\in\hr\otimes \hr$ being an arbitrary purification of the state $\rho$.\\
For the approximation of arbitrary compound channels by finite ones we use the diamond norm $||\cdot||_\lozenge$, which is given by
\[||\cn||_{\lozenge}:=\sup_{n\in \nn}\max_{a\in \mathcal{B}(\cc^n\otimes\hr),||a||_1=1}||(\textrm{id}_{n}\otimes \mathcal{N})(a)||_1,   \]
where $\textrm{id}_n:\mathcal{B}(\cc^n)\to \mathcal{B}(\cc^n)$ is the identity channel, and $\mathcal{N}:\mathcal{B}(\hr)\to \mathcal{B}(\kr)$ is any linear map, not necessarily completely positive. The merits of $||\cdot||_{\lozenge}$ are due to the following facts (cf. \cite{kitaev}). First, $||\cn||_{\lozenge}=1$ for all $\cn\in\mathcal{C}(\hr,\kr)$. Thus, $\mathcal{C}(\hr,\kr)\subset S_{\lozenge}$, where $S_{\lozenge}$ denotes the unit sphere of the normed space $(\mathcal{B}(\mathcal{B}(\hr),\mathcal{B}(\kr)),||\cdot||_{\lozenge} )$. Moreover, $||\cn_1\otimes \cn_2||_{\lozenge}=||\cn_1||_{\lozenge}||\cn_2||_{\lozenge}$ for arbitrary linear maps $\cn_1,\cn_2:\mathcal{B}(\hr)\to \mathcal{B}(\kr) $.\\
We further use the diamond norm to define the function $D_\lozenge(\cdot,\cdot)$ on $\{(\fri,\fri'):\fri,\fri'\subset\mathcal C(\hr,\kr)\}$, which is for $\fri,\fri'\subset\mathcal C(\hr,\kr)$ given by
$$D_\lozenge(\fri,\fri'):=\max\{\sup_{\cn\in\fri}\inf_{\cn'\in\fri'}||\cn-\cn'||_\lozenge,\sup_{\cn'\in\fri'}\inf_{\cn\in\fri}||\cn-\cn'||_\lozenge\}.$$
For $\fri\subset\mathcal C(\hr,\kr)$ let $\bar\fri$ denote the closure of $\fri$ in $||\cdot||_\lozenge$. Then $D_\lozenge$ defines a metric on $\{(\fri,\fri'):\fri,\fri'\subset\mathcal C(\hr,\kr),\ \fri=\bar\fri,\ \fri'=\bar{\fri'}\}$ which is basically the Hausdorff distance induced by the diamond norm.\\
Obviously, for arbitrary $\fri,\fri'\subset\mathcal C(\hr,\kr)$, $D_\lozenge(\fri,\fri')\leq\epsilon$ implies that for every $\cn\in\fri$ ($\cn'\in\fri'$) there exists $\cn'\in\fri'$ ($\cn\in\fri)$ such that $||\cn-\cn'||_\lozenge\leq2\epsilon$. If $\fri=\bar\fri,\ \fri'=\bar{\fri'}$ holds we even have $||\cn-\cn'||_\lozenge\leq\epsilon$.
In this way $D_\lozenge$ gives a measure of distance between two compound channels.\\
Finally, for any set $\fri\subset \mathcal{C}(\hr,\kr) $ and $l\in\nn$ we set
\[\fri^{\otimes l}:=\{\cn^{\otimes l}: \cn\in\fri  \}.  \]
\section{\label{sec:codes-and-capacity}Definitions and Main Result}
Let $\fri\subset \mathcal{C}(\hr,\kr)$. The memoryless compound channel associated with $\fri$ is given by the family $\{\cn^{\otimes l}:\mathcal{S}(\hr^{\otimes l})\to\mathcal{S}(\kr^{\otimes l})  \}_{\l\in\nn,\cn\in\fri}$. In the rest of the paper we will simply write $\fri$ for that family.\\
Each compound channel can be used in three different scenarios:
\begin{enumerate}
\item the \emph{informed decoder}
 \item the \emph{informed
encoder}
\item the case of \emph{uninformed users}.
\end{enumerate}
In the following three subsections we will give definitions of codes and capacity for these cases.
\subsection{\label{subsec:Definition of Codes:The Informed Decoder}The Informed Decoder}
An $(l,k_l)$-\emph{code} for $\fri$ with \emph{informed decoder}  is a pair $(\mathcal{P}^l,\{\crr^l_\cn:\cn\in\fri\})$ where:
\begin{enumerate}
\item $ \mathcal{P}^l:\mathcal{B}(\fr_l)\to\mathcal{B}(\hr)^{\otimes l} $ is a CPTP map for some Hilbert space $\fr_l$ with $k_l=\dim \fr_l$.
\item $\crr^l_\cn:\mathcal{B}(\kr)^{\otimes l}\to \mathcal{B}(\fr_l') $ is a CPTP map for each $\cn\in \fri$ where the Hilbert space $\fr_l'$ satisfies $\fr_l\subset \fr_l'$. In what follows the operations $\crr^l_\cn$ are referred to as recovery (or decoding) operations. Since the decoder knows which channel is actually used during transmission, they are allowed to depend on the channel.
\end{enumerate}
Note at this point that we deviate from the standard assumption that $\fr_l=\fr_l'$. We allow $\fr_l\subsetneq\fr_l'$ for convenience only since it allows more flexibility in code construction. It is readily seen from the definition of achievable rates and capacity below that the assumption $\fr_l\subsetneq\fr_l'$ cannot lead to a higher capacity of $\fri$ in any of the three cases that we are dealing with.\\
A non-negative number $R$ is called an achievable rate for
$\fri$ with informed decoder if there is a
sequence of $(l,k_l)$-codes such that
\begin{enumerate}
\item $\liminf_{l\to\infty}\frac{1}{l}\log k_l\ge R$, and \item
$\lim_{l\to\infty}\inf_{\cn\in\fri}F_e(\pi_{\fr_l},\crr^l_\cn\circ
\cn^{\otimes l}\circ\mathcal{P}^l )=1 $
\end{enumerate}
holds.\\
The \emph{capacity} $Q_{ID}(\fri)$ of the
compound channel $\fri$ with informed decoder
is given by
\begin{eqnarray*}
 Q_{ID}(\fri):=\sup\{R\in\rr_{+}&:& R \textrm{ is achievable for } \fri  \textrm{ with informed decoder}   \}.
\end{eqnarray*}
\subsection{\label{subsec:Definition of Codes:The Informed Encoder}The Informed Encoder}
An $(l,k_l)$-\emph{code} for $\fri$ with \emph{informed encoder}  is a pair $(\{\mathcal{P}_{\cn}^l: \cn\in \fri  \},\crr^l)$ where:
\begin{enumerate}
\item $ \mathcal{P}_{\cn}^l:\mathcal{B}(\fr_l)\to\mathcal{B}(\hr)^{\otimes l} $ is a CPTP map for each $\cn\in \fri$ for some Hilbert space $\fr_l$ with $k_l=\dim \fr_l$. The maps $\mathcal{P}_{\cn}^l$ are the encoding operations which we allow to depend on $\cn$ since the encoder knows which channel is in use.
\item $\crr^l:\mathcal{B}(\kr)^{\otimes l}\to \mathcal{B}(\fr_l') $ is a CPTP map where the Hilbert space $\fr_l'$ satisfies $\fr_l\subset \fr_l'$.
\end{enumerate}
A non-negative number $R$ is called an achievable rate for
$\fri$ with informed encoder if there is a
sequence of $(l,k_l)$-codes such that
\begin{enumerate}
\item $\liminf_{l\to\infty}\frac{1}{l}\log k_l\ge R$, and \item
$\lim_{l\to\infty}\inf_{\cn\in\fri}F_e(\pi_{\fr_l},\crr^l\circ
\cn^{\otimes l}\circ\mathcal{P}_{\cn}^l )=1 $
\end{enumerate}
holds.\\
The \emph{capacity} $Q_{IE}(\fri)$ of the
compound channel $\fri$ with informed encoder
is given by
\begin{eqnarray*}
 Q_{IE}(\fri):=\sup\{R\in\rr_{+}&:& R \textrm{ is achievable for } \fri  \textrm{ with informed encoder}   \}.
\end{eqnarray*}
\subsection{\label{subsec:Definition of Codes:The Case of Uninformed Users}The Case of Uninformed Users}
Codes and capacity for the compound channel $\fri$ with \emph{uninformed users} are defined in a similar fashion. The only change is that we do not allow the encoding operations to depend on $\cn$. I.e.  An $(l,k_l)-$ \emph{code} for $\fri$ is a pair $(\mathcal{P}^l, \crr^l)$ of CPTP maps $\mathcal{P}^l\in\mathcal{C}(\fr_l,\hr^{\otimes l})  $ where $\fr_l$ is a Hilbert space with $k_l=\dim \fr_l$ and $\crr^l\in\mathcal{C}(\kr^{\otimes l},\fr_l')$ with $\fr_l\subset \fr_l'$.\\
A non-negative number $R$ is called an achievable rate for
$\fri$ if there is a sequence of
$(l,k_l)$-codes such that
\begin{enumerate}
\item $\liminf_{l\to\infty}\frac{1}{l}\log k_l\ge R$, and \item
$\lim_{l\to\infty}\inf_{\cn\in\fri}F_e(\pi_{\fr_l},\crr^l\circ \cn^{\otimes l}\circ\mathcal{P}^l)=1 $.
\end{enumerate}
The \emph{capacity} $Q(\fri)$ of the compound
channel $\fri$ is given by
\begin{equation*}
 Q(\fri):=\sup\{R\in\rr_{+}: R \textrm{ is achievable for } \fri   \}.
\end{equation*}
A first simple consequence of these definitions is the following relation among the capacities of $\fri$.
\[ Q(\fri)\le \min \{ Q_{ID}(\fri),Q_{IE}(\fri) \}. \]
\subsection{\label{subsec:main-result} Main Result}

With these definitions at our disposal, we are ready now to state the main result of the paper. 
\begin{theorem}\label{main-theorem}
Let $\fri\subset\mathcal{C}(\hr,\kr)$ be an arbitrary set of quantum channels where $\hr$ and $\kr$ are  finite dimensional Hilbert spaces. 
\begin{enumerate}
\item Then
\[Q(\fri)=Q_{ID}(\fri)=  \lim_{l\to\infty}\frac{1}{l} \max_{\rho\in\mathcal{S}(\hr^{\otimes l})}\inf_{\cn\in\fri}I_c(\rho,\cn^{\otimes l}), \]
and
\[Q_{IE}(\fri)= \lim_{l\to\infty}\frac{1}{l}\inf_{\cn\in\fri} \max_{\rho\in\mathcal{S}(\hr^{\otimes l})}I_c(\rho,\cn^{\otimes l}).  \]
\item Moreover, for the corresponding entanglement-generating capacities $E(\fri)$, $E_{ID}(\fri)$, and $E_{IE}(\fri)$ (defined in Section \ref{sec:ent-generating}) we have
\[E(\fri)=E_{ID}(\fri)=Q(\fri)  \]
and
\[E_{IE}(\fri)=Q_{IE}(\fri).  \]
\end{enumerate}
\end{theorem}
The rest of the paper contains a step-by-step proof of Theorem \ref{main-theorem}.

\section{\label{sec:One-Shot Results}One-Shot Results}
In this section we will establish the basic building blocks for the achievability parts of the coding theorems for compound channels with and without channel knowledge. The results are formulated as one-shot statements in order to simplify the notation.

\subsection{\label{subsec:One-Shot Coding Result for Single Channel}One-Shot Coding Result for a Single Channel}
Before we turn our attention to quantum compound channels we will shortly describe a part of recent developments in coding theory for single (i.e. perfectly known) channels as given in \cite{klesse} and \cite{hayden}. Both approaches are based on a decoupling idea which is closely related to approximate error correction. In order to state this decoupling lemma we need some notational preparation.\\
Let $\rho\in\mathcal{S}(\hr)$ be given and consider any purification $\psi\in\hr_a\otimes \hr$, $\hr_a=\hr$, of $\rho$. According to Stinespring's representation theorem any $\cn\in\mathcal{C}^{\downarrow}(\hr,\kr)$ is given by
\begin{equation}\label{stinespring-1}
  \cn (\ \cdot\ )=\textrm{tr}_{\hr_e}((\idn_{\hr}\otimes p_e)v(\ \cdot\ )v^{\ast} ),
\end{equation}
where $\hr_e$ is a suitable finite-dimensional Hilbert space, $p_e$ is a projection onto a subspace of $\hr_e$, and $v:\hr \to\kr\otimes \hr_e $ is an isometry. \\
Let us define a pure state on $\hr_a\otimes \kr\otimes \hr_e$ by the formula
\[  \psi':= \frac{1}{\sqrt{\textrm{tr}(\cn(\pi_{\fr}))}}(\idn_{\hr_a\otimes \kr}\otimes p_e)(\idn_{\hr_a}\otimes v) \psi.  \]
We set
\[\rho':=\textrm{tr}_{\hr_a\otimes \hr_e}(|\psi'\rangle\langle \psi'|),\quad \rho'_{ae}:= \textrm{tr}_{\kr}(|\psi'\rangle\langle \psi'|), \]
and
\[\rho_a:=\textrm{tr}_{\kr\otimes\hr_e}(|\psi'\rangle\langle \psi'|),\quad \rho'_e:=\textrm{tr}_{\hr_a\otimes \kr}(|\psi'\rangle\langle \psi'|).  \]
The announced decoupling lemma can now be stated as follows.
\begin{lemma}[Cf. \cite{klesse},\cite{hayden}]\label{decoupling-lemma}
For $\rho\in\mathcal{S}(\hr)$ and $\cn\in \mathcal{C}^{\downarrow}(\hr,\kr) $ there exists a
recovery operation $\crr \in \mathcal{C}(\kr,\hr) $ with
\[F_e(\rho, \crr\circ \cn)\ge w-||w\rho'_{ae}-w\rho_a\otimes \rho'_e||_1,  \]
where $w=\textrm{tr}(\cn(\rho))$.
\end{lemma}
The striking implication of Lemma \ref{decoupling-lemma} is that if the so called quantum error $||\rho'_{ae}-\rho_a\otimes \rho'_e||_1 $ for $\rho\in\mathcal{S}(\hr)$ and $\cn\in\mathcal{C}(\hr,\kr)$ is small then almost perfect error correction is possible via $\crr$.\\
Lemma \ref{decoupling-lemma} was Klesse's  \cite{klesse} starting point for his highly interesting proof  of the following theorem which is a one-shot version of the achievability part of the coding theorem. In the statement of the result we will use the following notation.
\[F_{c,e}(\rho,\cn):=\max_{\crr \in \mathcal{C}(\kr,\hr)}F_e(\rho,\crr\circ\cn), \]
where $\rho\in\mathcal{S}(\hr)$ and $\cn\in\mathcal{C}^{\downarrow}(\hr,\kr) $.
\begin{theorem}[Klesse \cite{klesse}]\label{klesse-theorem}
Let the Hilbert space $\hr$ be given and consider subspaces
$\E\subset\gr\subset \hr$ with $\dim \E=k$. Then for any $\cn\in
\mathcal{C}^{\downarrow}(\hr,\kr)$ allowing a representation with
$n$ Kraus operators we have
\[\int_{\mathfrak{U}(\gr)}F_{c,e}(u\pi_{\E}u^{\ast},\cn)du\ge \textrm{tr}(\cn(\pi_{\gr}))-\sqrt{k\cdot n}||\cn(\pi_{\gr})||_2,  \]
where $\mathfrak{U}(\gr) $ denotes the group of unitaries acting
on $\gr$ and $du$ indicates that the integration is with respect
to the Haar measure on $\mathfrak{U}(\gr) $.
\end{theorem}
We will indicate briefly how Klesse \cite{klesse} derived the direct part of the coding theorem for memoryless quantum channels from Theorem \ref{klesse-theorem}.
Let us choose for each $l\in\nn$ subspaces $\E_l\subset \gr^{\otimes l}\subset \hr^{\otimes l}$ with
\[\dim \E_l=:k_l =2^{l(I_c(\pi_{\gr},\cn )-3\epsilon)}.   \]
To given $\cn\in\mathcal{C}(\hr,\kr)$ and $\pi_{\gr}$ Klesse constructed a reduced version $\cn_l$ of $\cn^{\otimes l}$ in such a way that $\cn_l$ has a Kraus representation with $n_l\le 2^{l(S_e(\pi_{\gr},\cn)+\epsilon   )} $ Kraus operators.
Let $q_l\in\mathcal{B}(\kr^{\otimes l})$ be the entropy-typical projection of the state $(\cn(\pi_{\gr}))^{\otimes l}$ and set $\cn'_l(\cdot):=q_l\cn_l(\cdot )q_l$. Then we have the following properties (some of which are stated once more for completeness)
\begin{enumerate}
\item $k_l=2^{l(I_c(\pi_{\gr},\cn )-3\epsilon)} $,
\item $\textrm{tr}(\cn'_l(\pi_{\gr}^{\otimes l}))\ge 1-o(l^0)\footnote{Here, $o(l^0)$ denotes simply a non-specified sequence tending to $0$ as $l\to\infty$, i.e. we (ab)use the Bachmann-Landau little-o notation.} $,
\item $n_l\le 2^{l(S_e(\pi_{\gr},\cn)+\epsilon   )} $, and
\item $||\cn'_l(\pi_{\gr}^{\otimes l})||_2^2\le 2^{-l(S(\pi_{\gr})-\epsilon   )} $
\end{enumerate}
An application of Theorem \ref{klesse-theorem} to $\cn'_l$ shows heuristically the existence of a unitary $u\in \mathfrak{U}(\gr^{\otimes l})$ and a recovery operation $\crr_l\in\mathcal{C}(\kr^{\otimes l},\hr^{\otimes l})$ with
\[ F_e(u\pi_{\E_l}u^{\ast},\crr_l\circ \cn'_l )\ge 1-o(l^0) -2^{-\frac{l}{2}\epsilon}. \]
This in turn can be converted into
\[ F_e(u\pi_{\E_l}u^{\ast},\crr_l\circ \cn^{\otimes l} )\ge 1-o(l^0), \]
which is the achievability of $I_c(\pi_{\gr},\cn)$. The passage from $\pi_{\gr}$ to arbitrary states $\rho$ is then accomplished via the Bennett, Shor, Smolin, and Thapliyal Lemma from \cite{bsst} and the rest is by regularization.

\subsection{\label{subsec:One-Shot Coding Result for Uninformed Users}One-Shot Coding Result for Uninformed Users}
Our goal in this section is to establish a variant of Theorem \ref{klesse-theorem} that works for finite sets of channels. Since the entanglement fidelity depends affinely on the channel it is easily seen that for each set $\fri=\{\cn_1,\ldots,\cn_N  \}$ any good coding scheme with uninformed users is also good for the channel
\[\cn:=\frac{1}{N}\sum_{i=1}^N\cn_i  \]
and vice versa. Since it is easier to deal with a single channel and we do not loose anything if passing to averages we will formulate our next theorem for arithmetic averages of completely positive trace decreasing maps instead of the set $\{\cn_1,\ldots,\cn_N  \}$.
\begin{theorem}[One-Shot Result: Uninformed Users and Averaged Channel]\label{convex-klesse}
Let the Hilbert space $\hr$ be given and consider subspaces
$\E\subset\gr\subset \hr$ with $\dim \E=k$. For any choice of
$\cn_1,\ldots \cn_N\in \mathcal{C}^{\downarrow}(\hr,\kr) $ each
allowing a representation with $n_j$ Kraus operators, $j=1,\ldots
, N$, we set
\[\cn:=\frac{1}{N}\sum_{j=1}^{N}\cn_j,  \]
and and for any $u\in\mathfrak{U}(\gr)$
\[\cn_u:= \frac{1}{N}\sum_{j=1}^{N}\cn_j\circ \U.   \]
Then
\begin{eqnarray*}
 \int_{\mathfrak{U}(\gr)}F_{c,e}(\pi_{\E},\cn_u)du&\ge& \textrm{tr}(\cn(\pi_{\gr}))
- 2 \sum_{j=1}^N \sqrt{k n_j }||\cn_j (\pi_{\gr})||_2,
\end{eqnarray*}
where the integration is with respect to the normalized Haar measure on $\mathfrak{U}(\gr)$.
\end{theorem}
\begin{remark}It is worth noting that the average in this theorem is no more over maximally mixed states like in Theorem \ref{klesse-theorem}, but rather over encoding operations.\end{remark}
\emph{Proof.} The proof is easily reduced to that of the corresponding theorem in our previous paper \cite{bbn-1}. Most of the details can also be seen in the proof of Theorem \ref{uie-theorem} in the next subsection.
\begin{flushright}$\Box$\end{flushright}
\subsection{\label{subsec:One-Shot Coding Result for Informed Encoder}One-Shot Coding Result for Informed Encoder}
Before stating the main result of this section we recall a useful lemma from \cite{bbn-1} which will be needed in the proof of Theorem \ref{uie-theorem}.
\begin{lemma}\label{matrix-lemma}
Let $L$ and $D$ be $N\times N$ matrices with non-negative entries
which satisfy
\begin{equation}\label{matrix-1}
 L_{jl}\le L_{jj}, \quad L_{j l}\le L_{ll},
\end{equation}
and
\begin{equation}\label{matrix-2}
 D_{jl}\le \max \{ D_{jj}, D_{ll} \}
\end{equation}
for all $j,l\in \{ 1,\ldots, N \}$. Then
\[\sum_{j,l=1}^{N}\frac{1}{N}\sqrt{L_{jl}D_{jl}}\le 2\sum_{j=1}^{N}\sqrt{L_{jj}D_{jj}}.  \]
\end{lemma}
\emph{Proof.} The proof of this lemma is elementary. The details
can be picked up in our previous paper \cite{bbn-1}.
\begin{flushright}$\Box$\end{flushright}
We will focus now on the scenario where the sender or encoder knows which channel is in use. Consequently, the encoding operation can depend on the individual channel. The idea behind the next theorem is that we perform an independent, randomized selection of unitary encoders for each channel in the finite set $\fri=\{\cn_1,\ldots,\cn_N  \}$. This explains why the averaging in (\ref{uie-inequality}) is with respect to products of Haar measures instead of averaging over one single Haar measure as in Theorem \ref{convex-klesse}.
\begin{theorem}[One-Shot Result: Informed Encoder and Averaged Channel]\label{uie-theorem}
Let the finite-dimensional Hilbert spaces $\hr$ and $\kr$ be
given. Consider subspaces $\E,\gr_1,\ldots,\gr_N\subset\hr$ with
$\dim \E=k$ such that for all $i\in\{1,\ldots,N\}$ the dimension relation $k\leq\dim\gr_i$
holds. Let $\cn_1,\ldots \cn_N\in
\mathcal{C}^{\downarrow}(\hr,\kr) $ each allowing a representation
with $n_j$ Kraus operators, $j=1,\ldots , N$. Let
$\{v_i\}_{i=1}^N\subset\mathfrak U(\hr)$ be any fixed set of
unitary operators such that $v_i\E \subset\gr_i$ holds for every
$i\in\{1,\ldots,N\}$. For an arbitrary set
$\{u_i\}_{i=1}^N\subset\mathfrak U(\hr)$, define
$$\cn_{u_1,\ldots,u_N}:=\frac{1}{N}\sum_{i=1}^N\cn_i\circ\U_i\circ\V_i.$$
Then
\begin{eqnarray}\label{uie-inequality}
\int_{\mathfrak{U}(\gr_1)\times\ldots
\times\mathfrak{U}(\gr_N)}F_{c,e}(\pi_{\E},\cn_{u_1,\ldots,u_N})du_1\ldots
du_N &\ge& \sum_{j=1}^N\Big[\frac{1}{N}\textrm{tr}(\cn_j(\pi_{\gr_j}))\nonumber\\
&& - 2\sqrt{k n_j }||\cn_j (\pi_{\gr_j})||_2\Big],
\end{eqnarray}
where the integration is with respect to the product of the
normalized Haar measures on $\mathfrak{U}(\gr_1),\ldots,
\mathfrak{U}(\gr_N)$.
\end{theorem}
\emph{Proof.} Our first step in the proof is to show briefly that
$F_{c,e}(\pi_{\E},\cn_{u_1,\ldots,u_N}) $ depends measurably on
$(u_1,\ldots ,u_N)\in \mathfrak{U}(\gr_1)\times\ldots \times
\mathfrak{U}(\gr_N)$. For each recovery operation
$\crr\in\mathcal{C}(\kr,\hr)$ we define a function
$f_{\crr}:\mathfrak{U}(\gr_1)\times \ldots\times
\mathfrak{U}(\gr_N)\to [0,1]$ by
\[f_{\crr}(u_1,\ldots,u_N):= F_e(\pi_{\E},\crr\circ \cn_{u_1,\ldots,u_N} ).  \]
Clearly, $f_{\crr}$ is continuous for each fixed
$\crr\in\mathcal{C}(\kr,\hr)$. Thus, the function
\[ F_{c,e}(\pi_{\E},\cn_{u_1,\ldots,u_N})=\max_{\crr\in\mathcal{C}(\kr,\hr)}f_{\crr}(u_1,\ldots,u_N) \]
is lower semicontinuous, and consequently measurable.\\
We turn now to the proof of inequality (\ref{uie-inequality}).
From Lemma \ref{decoupling-lemma} we know that there is a recovery
operation $\crr$ such that
\begin{equation}\label{eq:uie-theorem-1}
 F_e(\pi_{\E}, \crr\circ \cn_{u_1,\ldots,u_N})\ge w-||w\rho'_{ae}-w\rho_a\otimes \rho'_e||_1,
\end{equation}
where we have used the notation introduced in the paragraph
preceding Lemma \ref{decoupling-lemma}, and
\[ w=w(u_1,\ldots, u_N)=\textrm{tr}(\cn_{u_1,\ldots,u_N}(\pi_{\E} ) ). \]
For each $j\in\{1,\ldots ,N\}$ let $\{b_{j,i}\}_{i=1}^{n_j}$ be
the set of Kraus operators of $\cn_j$. Clearly, for every set
$u_1,\ldots,u_N$ of unitary matrices, $\cn_j\circ\U_j\circ\V_j$
has Kraus operators $\{a_{j,i}\}_{i=1}^{n_j}$ given by
$a_{j,i}=b_{j,i}u_jv_j$. Utilizing the very same calculation that
was used in the proof of Theorem \ref{convex-klesse} in
\cite{bbn-1}, which in turn is almost identical to the
corresponding calculation in \cite{klesse}, we can reformulate
inequality (\ref{eq:uie-theorem-1}) as
\begin{equation}\label{eq:uie-theorem-2}
   F_e(\pi_{\E}, \crr\circ \cn_{u_1,\ldots,u_N})\ge w-||D(u_1,\ldots,u_N)||_1,
\end{equation}
with $w=\textrm{tr}(\cn_{u_1,\ldots, u_N}(\pi_{\E}))$ and
\[D(u_1,\ldots,u_N):= \sum_{j,l=1}^{N}\frac{1}{N}\sum_{i,r=1}^{n_j,n_l}D_{(ij)(rl)}(u_j,u_l)  \otimes |e_i\rangle \langle e_r|\otimes |f_j\rangle\langle f_l|  \]
where
\[D_{(ij)(rl)}(u_j,u_l):= \frac{1}{k}\left(pa_{j,i}a_{l,r}^{\ast}p- \frac{1}{k}\textrm{tr}(pa_{j,i}^{\ast}a_{l,r}p )p\right) , \]
and $p:=k\pi_{\E}$ is the projection onto $\E$. Let us define
\begin{equation}\label{eq:uie-theorem-3}
D_{j,l}(u_j,u_l):=
\sum_{i=1,k=1}^{n_j,n_l}D_{(ij)(kl)}(u_j,u_l)\otimes|e_i\rangle
\langle e_k|\otimes |f_j\rangle\langle f_l|.
\end{equation}
The triangle inequality for the trace norm yields
\begin{eqnarray}\label{eq:uie-theorem-4}
  ||D(u_1,\ldots,u_N)||_1&\le& \sum_{j,l=1}^{N}\frac{1}{N}||D_{j,l}(u_j,u_l)||_1 \nonumber\\
&\le & \sum_{j,l=1}^{N}\frac{1}{N}\sqrt{k\min \{n_j,n_l  \}} ||D_{j,l}(u_j,u_l)||_2,\nonumber\\
&=& \sum_{j,l=1}^{N}\frac{1}{N}\sqrt{k\min \{n_j,n_l  \}
||D_{j,l}(u_j,u_l)||_2^2},
\end{eqnarray}
where the second line follows from $||a ||_1\le \sqrt{d}||a ||_2$, $d$ being the number of non-zero singular values of $a$.\\
In the next step we will compute $||D_{j,l}(u_j,u_l)||_2^2 $. We
set $p_l:=v_lpv_l^\ast$ which defines new projections
$\{p_l\}_{l=1}^N$ with $\supp (p_l)\subset\gr_l$ for every
$l\in\{1,\ldots,N\}$. A glance at (\ref{eq:uie-theorem-3}) shows
that
\begin{equation}\label{eq:uie-theorem-5}
  (D_{j,l}(u_j,u_l))^{\ast}=\sum_{i=1,k=1}^{n_j,n_l}(D_{(ij)(kl)}(u_j,u_l))^{\ast}\otimes|e_k\rangle \langle e_i|\otimes |f_l\rangle\langle f_j|,
\end{equation}
and consequently we obtain
\begin{eqnarray}\label{eq:uie-theorem-6}
||D_{j,l}(u_j,u_l)||_2^2&=& \textrm{tr}((D_{j,l}(u_j,u_l) )^{\ast}D_{j,l}(u_j,u_l) )  \nonumber\\
&=& \sum_{i=1,r=1}^{n_j,n_l}\textrm{tr}( (D_{(ij)(kl)}(u_j,u_l))^{\ast} D_{(ij)(kl)}(u_j,u_l))\nonumber\\
&=& \frac{1}{k^2}\sum_{i=1,r=1}^{n_j,n_l}\{\textrm{tr}(p (a_{j,i}^{\ast}a_{l,r})^{\ast}p a_{j,i}^{\ast}a_{l,r})\nonumber\\
& &-\frac{1}{k}|\textrm{tr}(pa_{j,i}^{\ast}a_{l,r} )|^2 \}\nonumber\\
&=& \frac{1}{k^2}\sum_{i=1,r=1}^{n_j,n_l}\{\textrm{tr}(p_l u_l^\ast b_{l,r}^{\ast}b_{j,i}u_jp_j u_j^\ast b_{j,i}^{\ast}b_{l,r}u_l)\nonumber\\
& &
-\frac{1}{k}|\textrm{tr}(pv_j^{\ast}u_j^{\ast}b_{j,i}^{\ast}b_{l,r}u_lv_l
)|^2 \}    .
\end{eqnarray}
It is apparent from the last two lines in (\ref{eq:uie-theorem-6})
that $||D_{j,l}(u_j,u_l)||_2^2 $ depends measurably on
$(u_1,\ldots ,u_N)\in \mathfrak{U}(\gr_1)\times\ldots \times
\mathfrak{U}(\gr_N) $. Let $U_1,\ldots, U_N$ be independent random
variables taking values in $\mathfrak{U}(\gr_i)$ according to the
normalized Haar measure on $\mathfrak{U}(\gr_i) $
($i\in\{1,\ldots,N\}$). Then using Jensen's inequality and
abbreviating $L_{jl}:=k\min\{n_j, n_l  \} $ we can infer from
(\ref{eq:uie-theorem-4}) that
\begin{equation}\label{eq:uie-theorem-7}
  \mathbb{E}( ||D(U_1,\ldots,U_N)||_1 )\le \sum_{j,l=1}^{N}\frac{1}{N}\sqrt{L_{jl}\mathbb{E}(||D_{j,l}(U_j,U_l) ||_2^2) }.
\end{equation}
Note that the expectations on the RHS of (\ref{eq:uie-theorem-7}) are only with respect to pairs of random variables $U_1,\ldots, U_N$.\\
Our next goal is to upper-bound $\mathbb{E}(||D_{j,l}(U_j,U_l) ||_2^2) $.\\
\emph{Case $j\neq l$:} Since the last term in
(\ref{eq:uie-theorem-6}) is non-negative and the random variables
$U_j$ and $U_l$ are independent we obtain the following chain of
inequalities:
\begin{eqnarray}\label{eq:uie-theorem-8}
\mathbb{E}(||D_{j,l}(U_j,U_l) ||_2^2)&=& \frac{1}{k^2}\sum_{i=1,r=1}^{n_j,n_l}\Big[\mathbb{E}\textrm{tr}(p_l U_l^\ast b_{l,r}^{\ast}b_{j,i}U_jp_j U_j^\ast b_{j,i}^{\ast}b_{l,r}U_l)\nonumber\\
& & -\frac{1}{k}\mathbb{E}|\textrm{tr}(pv_j^{\ast}U_j^{\ast}b_{j,i}^{\ast}b_{l,r}U_lv_l )|^2 \Big]\nonumber\\
&\le&  \frac{1}{k^2}\sum_{i=1,r=1}^{n_j,n_l}\mathbb{E}\textrm{tr}(p_l U_l^\ast b_{l,r}^{\ast}b_{j,i}U_jp_j U_j^\ast b_{j,i}^{\ast}b_{l,r}U_l)\nonumber\\
&=&  \frac{1}{k^2}\sum_{i=1,r=1}^{n_j,n_l}\mathbb{E}\textrm{tr}(U_l p_l U_l^\ast b_{l,r}^{\ast}b_{j,i}U_jp_j U_j^\ast b_{j,i}^{\ast}b_{l,r})\nonumber\\
&=&  \frac{1}{k^2}\sum_{i=1,r=1}^{n_j,n_l}\textrm{tr}(\mathbb{E}(U_lp_l U_l^\ast) b_{l,r}^{\ast}b_{j,i}\mathbb{E}(U_jp_j U_j^\ast) b_{j,i}^{\ast}b_{l,r})\nonumber\\
&=&\frac{1}{k^2}\sum_{i=1,r=1}^{n_j,n_l}\textrm{tr}(k\cdot\pi_{\gr_l} b_{l,r}^{\ast}b_{j,i}k\cdot\pi_{\gr_j} b_{j,i}^{\ast}b_{l,r})\nonumber\\
&=&\langle\cn_j(\pi_{\gr_j}),\cn_l(\pi_{\gr_l})\rangle_{HS},
\end{eqnarray}
where $\langle \ \cdot \ ,\ \cdot \ \rangle_{HS} $ denotes the
Hilbert-Schmidt inner product, and we used the fact that
\[ \mathbb{E}(U_lp_l U_l^\ast)= k\cdot\pi_{\gr_l}\quad\textrm{and}\quad \mathbb{E}(U_jp_j U_j^\ast)= k\cdot\pi_{\gr_j}. \]
\emph{Case $j=l$:}  In this case we obtain
\begin{eqnarray}\label{eq:uie-theorem-8a}
 \mathbb{E}(||D_{j,j}(U_j,U_j) ||_2^2)&=& \frac{1}{k^2}\sum_{i=1,r=1}^{n_j,n_j}\Big[\mathbb{E}\textrm{tr}(p_j U_j^\ast b_{j,r}^{\ast}b_{j,i}U_jp_j U_j^\ast b_{j,i}^{\ast}b_{j,r}U_j)\nonumber\\
& & -\frac{1}{k}\mathbb{E}|\textrm{tr}(pv_j^{\ast}U_j^{\ast}b_{j,i}^{\ast}b_{j,r}U_jv_j )|^2 \Big]\nonumber\\
&=& \frac{1}{k^2}\sum_{i=1,r=1}^{n_j,n_j}\mathbb{E}\textrm{tr}(U_j p_j U_j^\ast b_{j,r}^{\ast}b_{j,i}U_jp_j U_j^\ast b_{j,i}^{\ast}b_{j,r})\nonumber\\
& &
-\frac{1}{k}\mathbb{E}|\textrm{tr}(U_jp_jU_j^{\ast}b_{j,i}^{\ast}b_{j,r}
)|^2 \Big].
\end{eqnarray}
Thus, the problem reduces to the evaluation of
\[\mathbb{E}\{b_{UpU^{\ast}}(x,y)  \}, \qquad (x,y\in\mathcal{B}(\hr)) \]
where $p$ is an orthogonal projection with $\textrm{tr}(p)=k$ and
\[ b_{UpU^{\ast}}(x,y):= \textrm{tr}(UpU^{\ast} x^{\ast}UpU^{\ast}y )-\frac{1}{k}\textrm{tr}(UpU^{\ast}x^{\ast})\textrm{tr}(UpU^{\ast}y),  \]
for a Haar distributed random variable $U$ with values in $\mathfrak{U}(\gr)$ where $\supp(p)\subset\gr\subset\hr$.\\
Here we can refer to \cite{klesse} where the corresponding
calculation is carried out via the theory of group invariants and
explicit evaluations of appropriate integrals with respect to
row-distributions of random unitary matrices. The result is
\begin{equation}\label{eq:uie-theorem-8b}
  \mathbb{E}\{b_{UpU^{\ast}}(x,y) \}= \frac{k^2-1}{d^2-1}\textrm{tr}(p_{\gr}x^{\ast}p_{\gr}y)+\frac{1-k^2}{d(d^2-1 )}\textrm{tr}(p_{\gr}x^{\ast})\textrm{tr}(p_{\gr}y),
\end{equation}
for all $x,y\in\mathcal{B}(\hr)$ where $p_{\gr}$ denotes the projection onto $\gr$ with $\textrm{tr}(p_{\gr})=d$. In Appendix \ref{sec:appendix} we will give an elementary derivation of (\ref{eq:uie-theorem-8b}) for the sake of completeness.\\
Inserting (\ref{eq:uie-theorem-8b}) with
$x=y=b_{j,i}^{\ast}b_{j,r}$ into (\ref{eq:uie-theorem-8a}) yields
with $d_j:=\textrm{tr}(p_{\gr_{j}})$
\begin{eqnarray*}
   \mathbb{E}(||D_{j,j}(U_j,U_j) ||_2^2)&=&\frac{1-\frac{1}{k^2}}{d_j^2-1}\Big[\sum_{i=1,r=1}^{n_j,n_j}\textrm{tr}(p_{\gr_{j}} b_{j,r}^{\ast}b_{j,i}p_{\gr_{j}} b_{j,i}^{\ast}b_{j,r})\\
& &-\frac{1}{d_j}|\textrm{tr}((p_{\gr_{j}} b_{j,i}^{\ast}b_{j,r} )|^2 \Big]\\
&\le& \frac{1-\frac{1}{k^2}}{d_j^2-1}\sum_{i=1,r=1}^{n_j,n_j}\textrm{tr}(p_{\gr_{j}} b_{j,r}^{\ast}b_{j,i}p_{\gr_{j}} b_{j,i}^{\ast}b_{j,r})\\
&\le& \frac{1}{d_j^2}\sum_{i=1,r=1}^{n_j,n_j}\textrm{tr}(p_{\gr_{j}} b_{j,r}^{\ast}b_{j,i}p_{\gr_{j}} b_{j,i}^{\ast}b_{j,r})\\
&=&  \frac{1}{d_j^2}\sum_{i=1,r=1}^{n_j,n_j}\textrm{tr}(b_{j,r}p_{\gr_{j}} b_{j,r}^{\ast}b_{j,i}p_{\gr_{j}} b_{j,i}^{\ast})\\
&=& \langle \cn_j(\pi_{\gr_{j}} ),\cn_j(\pi_{\gr_{j}}
)\rangle_{HS}.
\end{eqnarray*}
Summarizing, we obtain
\begin{equation}\label{eq:uie-theorem-9}
 \mathbb{E}(||D_{j,j}(U_j,U_j) ||_2^2) \le \langle\cn_j(\pi_{\gr_j}),\cn_j(\pi_{\gr_j})\rangle_{HS}=||\cn_j(\pi_{\gr_j})||_2^2.
\end{equation}
Similarly
\begin{eqnarray}
\mathbb{E}(\textrm{tr}(\cn_{U_1,\ldots,U_N}(\pi_{\E})))&=&\frac{1}{N}\sum_{j=1}^N\mathbb E(\tr(\cn_j(U_j\frac{1}{k}p_jU_j^\ast)))\nonumber\\
&=&\frac{1}{N}\sum_{j=1}^N\textrm{tr}(\cn_j(\pi_{\gr_j})).\label{eq:uie-theorem-9a}
\end{eqnarray}
(\ref{eq:uie-theorem-2}), (\ref{eq:uie-theorem-4}),
(\ref{eq:uie-theorem-8}), (\ref{eq:uie-theorem-9}), and
(\ref{eq:uie-theorem-9a}) show that
\begin{eqnarray}\label{eq:uie-theorem-10}
 \mathbb{E} (F_{c,e}(\pi_{\E},\cn_{U_1,\ldots,U_N}))&\ge& \frac{1}{N}\sum_{j=1}^N\textrm{tr}(\cn_j (\pi_{\gr_j}))\nonumber\\
&& -\sum_{j,l=1}^{N}\frac{1}{N}\sqrt{L_{jl}D_{jl}},
\end{eqnarray}
where for $j,l\in\{1,\ldots, N  \}$ we introduced the abbreviation
\[ D_{jl}:= \langle\cn_j(\pi_{\gr_j}),\cn_l(\pi_{\gr_l})\rangle_{HS}, \]
and, as before,
\[L_{jl}=k\min\{n_j, n_l  \}.  \]
It is obvious that
\[  L_{jl}\le L_{jj} \quad \textrm{and}\quad  L_{j l}\le L_{ll}   \]
hold. Moreover, the Cauchy-Schwarz inequality for the
Hilbert-Schmidt inner product shows that
\begin{eqnarray*}
 D_{jl} &=& \langle\cn_j(\pi_{\gr_j}),\cn_l(\pi_{\gr_l})\rangle_{HS}\\
&\le & ||\cn_j(\pi_{\gr_j})||_2 ||\cn_l(\pi_{\gr_l})||_2\\
&\le & \max\{||\cn_j(\pi_{\gr_j})||_2^2, ||\cn_l(\pi_{\gr_l})||_2^2  \}\\
&=& \max\{D_{jj}, D_{ll}\} .
\end{eqnarray*}
 Therefore, an application of Lemma \ref{matrix-lemma} allows us to conclude from (\ref{eq:uie-theorem-10}) that
\begin{eqnarray*}
\mathbb{E} (F_{c,e}(\pi_{\E},\cn_{U_1,\ldots,U_N}))&\ge& \frac{1}{N}\sum_{j=1}^N\textrm{tr}(\cn_j (\pi_{\gr_j}))\nonumber\\
&& -2\sum_{j=1}^{N}\sqrt{kn_j}||\cn_j(\pi_{\gr_j})||_2,
\end{eqnarray*}
and we are done.
\begin{flushright}$\Box$\end{flushright}
\subsection{\label{subsec:Entanglement Fidelity}Entanglement Fidelity}
The purpose of this subsection is to develop a tool which will enable us to convert a special kind of recovery maps depending on the channel into such that are universal, at least for finite compound channels. Anticipating constructions in section \ref{sec:Direct Part of The Coding Theorem for Finitely many Channels} below the situation we will be faced with is as follows. For finite set $\fri=\{\cn_1,\ldots,\cn_N  \}$ of channels, block length $l\in \nn$, and small $\epsilon >0$ we will be able to find one single recovery map $\crr^l$ and a unitary encoder $\mathcal{W}^l$ such that for each $i\in\{1,\ldots, N  \}$
\[ F_e(\pi_{\fr_l},\crr^l\circ \mathcal{Q}_{l,i}\circ \cn_i^{\otimes l}\circ \mathcal{W}^l)\ge 1-\epsilon, \]
where $\mathcal{Q}_{l,i}(\cdot):=q_{l,i}(\cdot)q_{l,i} $ with suitable projections $q_{l,i}$ acting on $\kr^{\otimes l}$. Thus we will effectively end up with the recovery maps $\crr_i^l:=\crr^l\circ \mathcal{Q}_{l,i} $. Consequently, it turns out that the decoder is \emph{informed}. Lemma \ref{gentle-operator-lemma-for-F_e} below shows how to get rid of the maps $\mathcal{Q}_{l,i}$ ensuring the existence of a universal recovery map for the whole set $\fri$ while decreasing the entanglement fidelity only slightly.
\begin{lemma}\label{gentle-operator-lemma-for-F_e}
Let $\rho\in\mathcal S(\hr)$ for some Hilbert space $\hr$. Let, for some other Hilbert space $\kr$, $\A\in \mathcal C(\hr,\kr),\ \D\in \mathcal C(\kr,\hr)$, $q\in\mathcal B(\kr)$ be an orthogonal projection.
\begin{enumerate}
\item Denoting by $\mathcal{Q}^{\perp}$ the completely positive map induced by $q^{\perp}:=\idn_{\kr}-q$ we have
\begin{equation}\label{eq:F_e-lemma-1}
F_e(\rho,\D\circ\A)\geq F_e(\rho,\D\circ \mathcal{Q}\circ\A)(1-2 F_e(\rho,\D\circ \mathcal{Q}^{\perp}\circ\A) ).
\end{equation}
\item If for some $\epsilon>0$ the relation $F_e(\rho,\D\circ\mathcal{Q}\circ\A)\geq1-\epsilon $ holds, then
\[ F_e(\rho,\D\circ \mathcal{Q}^{\perp}\circ\A)\le \epsilon, \]
and (\ref{eq:F_e-lemma-1}) implies
\begin{equation}\label{eq:F_e-lemma-2}
F_e(\rho,\D\circ\A)\geq(1-\epsilon)(1-2\epsilon)\ge 1-3\epsilon.
\end{equation}
\item If for some $\epsilon>0$ merely the relation $\textrm{tr}\{q\A(\rho)\}\geq1-\epsilon$ holds then we can conclude that
  \begin{equation}\label{eq:F_e-lemma-3}
 F_e(\rho,\D\circ\A)\geq F_e(\rho,\D\circ \mathcal{Q}\circ\A)-2\epsilon.
  \end{equation}
\end{enumerate}

\end{lemma}
The following Lemma \ref{choi-lemma} contains two inequalities one of which will be needed in the proof of Lemma \ref{gentle-operator-lemma-for-F_e}.
\begin{lemma}\label{choi-lemma}
Let $\D\in\mathcal C(\kr,\hr)$ and $x_1\perp x_2$, $z_1\perp z_2$
be state vectors, $x_1,x_2\in\kr,\ z_1,z_2\in\hr$. Then
\begin{equation}\label{eq:choi-lemma-1}
|\langle z_1,\D(|x_1\rangle\langle x_2|)z_1\rangle|\leq\sqrt{|\langle z_1,\D(|x_1\rangle\langle x_1|)z_1\rangle|\cdot|\langle z_1,\D(|x_2\rangle\langle x_2|)z_1\rangle|}\leq1,
\end{equation}
and
\begin{equation}\label{eq:choi-lemma-2}
|\langle z_1,\D(|x_1\rangle\langle x_2|)z_2\rangle|\leq\sqrt{|\langle z_1,\D(|x_1\rangle\langle x_1|)z_1\rangle|\cdot|\langle z_2,\D(|x_2\rangle\langle x_2|)z_2\rangle|}\leq1.
\end{equation}
\end{lemma}
We will utilize only (\ref{eq:choi-lemma-1}) in the proof of Lemma \ref{gentle-operator-lemma-for-F_e}. But the inequality (\ref{eq:choi-lemma-2}) might prove useful in other context so that we state it here for completeness.

\emph{Proof of Lemma \ref{choi-lemma}.} Let $\dim\hr=h,\ \dim\kr=\kappa$. Extend $\{x_1,x_2\}$ to
an orthonormal basis $\{x_1,x_2,\ldots,x_\kappa\}$  of $\kr$ and
$\{z_1,z_2\}$ to an orthonormal basis $\{z_1,z_2,\ldots,z_h\}$ on
$\hr$. Since $x_1\perp x_2$ and $z_1\perp z_2$, this can always be
done. By the theorem of Choi \cite{choi}, a linear map from
$\mathcal B(\hr)$ to $\mathcal B(\kr)$ is completely positive if
and only if its Choi matrix is positive. Write
$\D(|x_i\rangle\langle
x_j|)=\sum_{k,l=1}^hD^{ij}_{kl}|z_k\rangle\langle z_l|$. Then the
Choi matrix of $\D$ is, with respect to the bases
$\{x_1,\ldots,x_k\}$ and $\{z_1,\ldots,z_h\}$, written as
$$\textrm{CHOI}(\D)=\sum_{i,j=1}^{\kappa}|x_i\rangle\langle x_j|\otimes\sum_{k,l=1}^hD^{ij}_{kl}|z_k\rangle\langle z_l|.$$
If $\textrm{CHOI}(\D)$ is positive, then all principal minors of $\textrm{CHOI}(\D)$ are positive (cf. Corollary 7.1.5 in \cite{horn-johnson}) and thus
$$|D^{ij}_{kl}|\leq\sqrt{|D^{ii}_{kk}|\cdot|D^{jj}_{ll}|}$$
for every suitable choice of $i,j,k,l$. Thus
\begin{eqnarray*}
|\langle z_1|\D(|x_1\rangle\langle x_2|)z_2\rangle|&=&|D^{12}_{12}|\\
&\leq&\sqrt{|D^{11}_{11}|\cdot|D^{22}_{22}|}\\
&=&\sqrt{|\langle z_1,\D(|x_1\rangle\langle
x_1|)z_1\rangle|\cdot|\langle z_2,\D(|x_2\rangle\langle
x_2|)z_2\rangle|},
\end{eqnarray*}
and similarly
\[|\langle z_1,\D(|x_1\rangle\langle x_2|)z_1\rangle|\leq\sqrt{|\langle z_1,\D(|x_1\rangle\langle x_1|)z_1\rangle|\cdot|\langle z_1,\D(|x_2\rangle\langle x_2|)z_1\rangle|}.\]
The fact that $\D$ is trace preserving gives us the estimate
$\langle z_i,\D(|x_j\rangle\langle x_j|)z_i\rangle\leq1$ ($i,j$
suitably chosen) and we are done.
\begin{flushright}$\Box$\end{flushright}

\emph{Proof of Lemma \ref{gentle-operator-lemma-for-F_e}.} Let $\dim\hr=h,\ \dim\kr=\kappa$,
$|\psi\rangle\langle\psi|\in\hr_a\otimes\hr$ be a purification of
$\rho$ (w.l.o.g. $\hr_a=\hr$). Set
$\tilde\D:=id_{\hr_a}\otimes\D,\ \tilde\A:=id_{\hr_a}\otimes\A,\
\tilde q:=\idn_{\hr_a}\otimes q$ and, as usual, $\tilde
q^\perp$ the orthocomplement of $\tilde q$ within
$\hr_a\otimes\kr$. Obviously,
\begin{eqnarray}
F_e(\rho,\D\circ\A)&=&\langle\psi,\tilde\D\circ\tilde\A(|\psi\rangle\langle\psi|)\psi\rangle\nonumber\\
&=&\langle\psi,\tilde\D([\tilde q+\tilde q^\perp]\tilde\A(|\psi\rangle\langle\psi|[\tilde q+\tilde q^\perp]))\psi\rangle\nonumber\\
&=&\langle\psi,\tilde\D(\tilde q \tilde\A(|\psi\rangle\langle\psi|)\tilde q)\psi\rangle+\langle\psi,\tilde\D(\tilde q^\perp\tilde\A(|\psi\rangle\langle\psi|)\tilde q^\perp)\psi\rangle\nonumber\\
&&+\langle\psi,\tilde\D(\tilde q\tilde\A(|\psi\rangle\langle\psi|)\tilde q^\perp)\psi\rangle+\langle\psi,\tilde\D(\tilde q^\perp\tilde\A(|\psi\rangle\langle\psi|)\tilde q)\psi\rangle\nonumber\\
&\geq&\langle\psi,\tilde\D(\tilde q \tilde\A(|\psi\rangle\langle\psi|)\tilde q)\psi\rangle+2\Re\{\langle\psi,\tilde\D(\tilde q\tilde\A(|\psi\rangle\langle\psi|)\tilde q^\perp)\psi\rangle\}\nonumber\\
&\geq&\langle\psi,\tilde\D(\tilde q \tilde\A(|\psi\rangle\langle\psi|)\tilde q)\psi\rangle-2|\langle\psi,\tilde\D(\tilde q\tilde\A(|\psi\rangle\langle\psi|)\tilde q^\perp)\psi\rangle|\nonumber\\
&=&F_e(\rho,\D\circ\mathcal
Q\circ\A)-2|\langle\psi,\tilde\D(\tilde
q\tilde\A(|\psi\rangle\langle\psi|)\tilde
q^\perp)\psi\rangle|\label{eq:Entanglement Fidelity-1}.
\end{eqnarray}
We establish a lower bound on the second term on the RHS of
(\ref{eq:Entanglement Fidelity-1}). Let
\[ \tilde\A(|\psi\rangle\langle\psi|)=\sum_{i=1}^{\kappa\cdot
h}\lambda_i|a_i\rangle\langle a_i|,\]
 where $\{a_1,\ldots,a_{\kappa\cdot
h}\}$ are assumed to form an orthonormal basis. Now every $a_i$
can be written as $a_i=\alpha_ix_i+\beta_iy_i$ where $x_i\in
\mathrm{supp}(\tilde q)$ and $y_i\in \mathrm{supp}(\tilde
q^\perp)$, $i\in\{1,...,\kappa\cdot h\}$, are state vectors and
$\alpha_i,\beta_i\in\mathbb C$. Define $\sigma:=\tilde
A(|\psi\rangle\langle\psi|)$, then
\begin{eqnarray}
\sigma=\sum_{j=1}^{\kappa\cdot
h}\lambda_j(|\alpha_j|^2|x_j\rangle\langle
x_j|+\alpha_j\beta_j^\ast|x_j\rangle\langle
y_j|+\beta_j\alpha_j^\ast|y_j\rangle\langle
x_j|+|\beta_j|^2|y_j\rangle\langle y_j|).\label{eq:Entanglement
Fidelity-2}
\end{eqnarray}
Set $X:=|\langle\psi,\tilde\D(\tilde
q\tilde\A(|\psi\rangle\langle\psi|)\tilde q^\perp)\psi\rangle|$.
Then
\begin{eqnarray}
X&=&|\langle\psi,\tilde\D(\tilde q\sigma q^\perp)\psi\rangle|\nonumber\\
&\overset{\mathbf a}{=}&|\sum_{i=1}^{\kappa\cdot h}\lambda_i\langle\psi,\tilde\D(\tilde q|a_i\rangle\langle a_i|\tilde q^\perp)\psi\rangle|\nonumber\\
&=&|\sum_{i=1}^{\kappa\cdot h}\lambda_i\alpha_i\beta_i^*\langle\psi,\tilde\D(|x_i\rangle\langle y_i|)\psi\rangle|\nonumber\\
&\leq&\sum_{i=1}^{\kappa\cdot h}|\lambda_i\alpha_i\beta_i^*|\cdot|\langle\psi,\tilde\D(|x_i\rangle\langle y_i|)\psi\rangle|\nonumber\\
&\overset{\mathbf b}{\leq}&\sum_{i=1}^{\kappa\cdot h}|\sqrt{\lambda_i|\langle\psi,\tilde\D(|x_i\rangle\langle x_i|)\psi\rangle}\alpha_i\sqrt{\lambda_i\langle\psi,\tilde\D(|y_i\rangle\langle y_i|)\psi\rangle}\beta_i^*|\nonumber\\
&\overset{\mathbf c}{\leq}&\sum_{i=1}^{\kappa\cdot
h}\lambda_i|\alpha_i|^2\langle\psi,\tilde\D(|x_i\rangle\langle
x_i|)\psi\rangle\sum_{j=1}^{\kappa\cdot
h}\lambda_j|\beta_j|^2\langle\psi,\tilde\D(|y_j\rangle\langle
y_j|)\psi\rangle.\label{eq:Entanglement Fidelity-4}
\end{eqnarray}
Here, $\mathbf a$ follows from using the convex decomposition of $\tilde\A(|\psi\rangle\langle\psi|)$, $\mathbf b$ from utilizing inequality (\ref{eq:choi-lemma-1}) from Lemma \ref{choi-lemma} and $\mathbf c$ is an application of the Cauchy-Schwarz inequality.\\
Now, employing the representation (\ref{eq:Entanglement Fidelity-2}) it is easily seen that
\begin{equation}\label{eq: Entanglement Fidelity-5}
   F_e(\rho,\D\circ\mathcal Q\circ\A)= \langle\psi,\tilde\D(\tilde q\tilde\A(|\psi\rangle\langle\psi|)\tilde q)\psi\rangle= \sum_{i=1}^{\kappa\cdot
h}\lambda_i|\alpha_i|^2\langle\psi,\tilde\D(|x_i\rangle\langle,
x_i|)\psi\rangle
\end{equation}
and similarly
\begin{equation}\label{eq:Entanglement Fidelity-6}
 F_e(\rho,\D\circ\mathcal Q\circ\A)= \sum_{j=1}^{\kappa\cdot
h}\lambda_j|\beta_j|^2\langle\psi,\tilde\D(|y_j\rangle\langle
y_j|)\psi\rangle.
\end{equation}
The inequalities (\ref{eq:Entanglement Fidelity-6}), (\ref{eq: Entanglement Fidelity-5}), (\ref{eq:Entanglement Fidelity-4}), and (\ref{eq:Entanglement Fidelity-1}) yield
\begin{eqnarray}\label{eq:Entanglement Fidelity-7}
F_e(\rho,\D\circ\A)&\ge& F_e(\rho,\D\circ\mathcal Q\circ\A)-2 F_e(\rho,\D\circ\mathcal Q\circ\A)F_e(\rho,\D\circ\mathcal{Q}^{\perp}\circ\A)\nonumber\\
&=&  F_e(\rho,\D\circ\mathcal Q\circ\A)(1-2 F_e(\rho,\D\circ\mathcal{Q}^{\perp}\circ\A) )
\end{eqnarray}
which establishes (\ref{eq:F_e-lemma-1}).\\
Let us turn now to the other assertions stated in the lemma. Let
$\textrm{tr}\{q\A(\rho)\}\geq 1-\epsilon$. This implies
$\tr(q^\perp\A(\rho))\leq\epsilon$. A direct calculation yields
\begin{eqnarray*}
\mathrm{tr}(\tilde q^\perp\sigma)&=&\mathrm{tr}_{\hr_a}(\mathrm{tr}_\kr((\mathbf{1}_{\hr_a}\otimes q^\perp)id_{\hr_a}\otimes\A(|\psi\rangle\langle\psi|)))\\
&=&\mathrm{tr}_\kr(q^\perp\A(\mathrm{tr}_{\hr_a}(|\psi\rangle\langle\psi|)))\\
&=&\mathrm{tr}_\kr(q^\perp\A(\rho))\\
&\leq&\epsilon.
\end{eqnarray*}
Using (\ref{eq:Entanglement Fidelity-2}), we get the useful
inequality
\begin{eqnarray}
\epsilon&\geq&\mathrm{tr}(\tilde q^\perp\sigma)\nonumber\\
&=&\sum_{i=1}^{\kappa\cdot h}\lambda_i|\beta_i|^2\mathrm{tr}(\tilde q^\perp|y_i\rangle\langle y_i|)\nonumber\\
&=&\sum_{i=1}^{\kappa\cdot
h}\lambda_i|\beta_i|^2.\label{eq:Entanglement Fidelity-3}
\end{eqnarray}
Using Lemma \ref{choi-lemma} and (\ref{eq:Entanglement
Fidelity-3}) we get
\begin{eqnarray*}
X&\leq&\sum_{i=1}^{\kappa\cdot h}\lambda_i|\alpha_i|^2\sum_{j=1}^{\kappa\cdot h}\lambda_j|\beta_j|^2\\
&\leq&\epsilon,
\end{eqnarray*}
thus by equation (\ref{eq:Entanglement Fidelity-1}) we have
$$F_e(\rho,\D\circ\A)\geq F_e(\rho,\D\circ\mathcal Q\circ\A)-2\epsilon.$$
In case that $F_e(\rho,\D\circ\mathcal Q\circ\A)\geq1-\epsilon$,
we note that the linear maps $\mathcal Q$ and $\mathcal Q^\perp$
are elements of $\mathcal C^\downarrow(\kr,\kr)$ whilst $\mathcal
Q+\mathcal Q^\perp\in\mathcal C(\kr,\kr)$ and since $F_e$ is
affine in the operation
\[F_e(\rho,\D\circ\mathcal Q\circ\A)+F_e(\rho,\D\circ\mathcal Q^\perp\circ\A)=F_e(\rho,\D\circ(\mathcal Q+\mathcal{Q}^{\perp})\circ\A  )\leq 1\]
has to
hold. This in turn implies
$$F_e(\rho,\D\circ\mathcal Q^\perp\circ\A)\leq\epsilon.$$
Using this, our assumption that $F_e(\rho,\D\circ\mathcal Q\circ\A)\geq1-\epsilon$, and  (\ref{eq:Entanglement Fidelity-7}) we obtain that
\begin{eqnarray*}
F_e(\rho,\D\circ\A)&\ge& F_e(\rho,\D\circ\mathcal Q\circ\A)(1-2 F_e(\rho,\D\circ\mathcal{Q}^{\perp}\circ\A))\\
&\ge& (1-\epsilon)(1-2\epsilon)\\
&\ge& 1-3\epsilon,
\end{eqnarray*}
which is the claim we made in (\ref{eq:F_e-lemma-2}).
\begin{flushright}$\Box$\end{flushright}

\section{\label{sec:Direct Part of The Coding Theorem for Finitely many Channels}Direct Part of The Coding Theorem for Finitely Many Channels}

\subsection{\label{subsec:Typical Projections and Kraus Operators}Typical Projections and Kraus Operators}
In this subsection we recall briefly the well-known properties
of frequency typical projections and reduced operations. A more
detailed description can be found in \cite{bbn-1} and references therein.
\begin{lemma}\label{lemma-typical-1}
There is a real number $c>0$ such that for every Hilbert space $\hr$ there exist functions $h:\mathbb
N\rightarrow\mathbb R_+$, $\varphi:(0,1/2)\rightarrow\mathbb R_+$
with $\lim_{l\rightarrow\infty}h(l)=0$ and $\lim_{\delta\to
0}\varphi(\delta)= 0$ such that for any $\rho\in \cs (\hr),\ \delta\in(0,1/2),\
l\in\mathbb N$ there is an orthogonal projection $q_{\delta,l}\in
\mathcal{B}(\hr)^{\otimes l}$ called frequency-typical projection
that satisfies
\begin{enumerate}
\item $\textrm{tr}(\rho^{\otimes l}q_{\delta,l})\ge
1-2^{-l(c\delta^2-h(l))}$,
\item $q_{\delta,l}\rho^{\otimes l}q_{\delta,l}\le
2^{-l(S(\rho)-\varphi(\delta) )}q_{\delta,l}  $.
\end{enumerate}
The inequality 2. implies
\[||q_{\delta,l}\rho^{\otimes l}q_{\delta,l}||_2^2 \le 2^{-l(S(\rho)-\varphi(\delta))}. \]
Moreover, setting $d:=\dim\hr$, $\vphi$ and $h$ are given by
\[h(l)=\frac{d}{l}\log(l+1)\ \ \forall l\in\mathbb N,\ \ \ \vphi(\delta)=-\delta\log \frac{\delta}{d}\ \forall\delta\in(0,1/2). \]
\end{lemma}
\begin{lemma}\label{lemma-typical-2}
Let $\hr,\kr$ be finite dimensional Hilbert spaces. There are functions
$\gamma:(0,1/2)\rightarrow\mathbb R_+$, $h':\mathbb N\rightarrow\mathbb R_+$ satisfying $\lim_{\delta\to
0}\gamma(\delta)=0$ and $h'(l)\searrow0$ such that for each $\cn\in
\mathcal{C}(\hr,\kr)$, $\delta\in (0,1/2)$, $l\in\nn$ and maximally mixed state $\pi_\gr$ on some
subspace $\gr\subset\hr$ there is an operation
$\cn_{\delta,l}\in\mathcal C^\downarrow(\hr^{\otimes
l},\kr^{\otimes l})$ called reduced operation with respect to
$\cn$ and $\pi_\gr$ that satisfies
\begin{enumerate}
\item $\textrm{tr}(\cn_{\delta,l}(\pi_{\gr}^{\otimes l}))\ge
1-2^{-l(c'\delta^2-h'(l))}$, with a universal positive constant
$c'>0$,
\item $\cn_{\delta,l}$ has a Kraus representation with at
most $n_{\delta,l}\le 2^{l(S_{e}(\pi_{\gr},\cn)+\gamma(\delta)+h'(l))}$
Kraus operators. \item For every state $\rho\in\mathcal
S(\hr^{\otimes l})$ and every two channels $\mathcal I\in\mathcal
C^\downarrow(\hr^{\otimes l},\hr^{\otimes l})$ and $\mathcal
L\in\mathcal C^\downarrow(\kr^{\otimes l},\hr^{\otimes l})$ the
inequality $F_e(\rho,\mathcal L\circ\cn_{\delta,l}\circ\mathcal
I)\leq F_e(\rho,\mathcal L\circ\cn^{\otimes l}\circ\mathcal I)$ is
fulfilled.
\end{enumerate}
Setting $d:=\dim\hr$ and $\kappa:=\dim\kr$, the function $h':\mathbb N\rightarrow\mathbb R_+$ is given by
$h'(l)=\frac{d\cdot\kappa}{l}\log(l+1)\ \forall l\in\mathbb N$ and $\gamma$ by $\gamma(\delta)=-\delta\log\frac{\delta}{d\cdot\kappa},\ \forall\delta\in(0,1/2)$.
\end{lemma}
\subsection{\label{subsec:The Case of Uninformed Users}The Case of Uninformed Users}
Let us consider a compound channel given by a finite set $\fri:=\{\cn_1,\ldots ,\cn_{N} \}\subset \mathcal{C}(\hr,\kr)$ and a subspace $\gr\subset\hr$. For every $l\in\mathbb N$, we choose a subspace $\E_l\subset\gr^{\otimes l}$. As usual, $\pi_{\E_l}$ and $\pi_\gr$ denote the maximally mixed states on $\E_l$, respectively $\gr$ while $k_l:=\dim\E_l$ gives the dimension of $\E_l$.\\
For $j\in\{1,\ldots,N\}$, $\delta\in(0,1/2)$, $l\in\nn$ and states $\cn_j(\pi_{\gr})$ let $q_{j,\delta,l}\in \mathcal{B}(\kr)^{\otimes l}$ be the frequency-typical projection of $\cn_j(\pi_\gr)$ and $\cn_{j,\delta,l}$ be the reduced operation associated with $\cn_j$ and $\pi_\gr$ as defined in Subsec. \ref{subsec:Typical Projections and Kraus Operators}.\\
These quantities enable us to define a new set of channels that is
more adapted to our problem than the original one. We set for an arbitrary unitary operation $u^l\in\mathcal B(\hr^{\otimes l})$
\begin{eqnarray*}
\hat\cn_{j,u^l,\delta}^l:=\mathcal{Q}_{j,\delta,l}\circ\cn_{j,\delta,l}\circ\mathcal U^l
\end{eqnarray*}
and, accordingly,
\[\hat\cn^l_{u^l,\delta}:=\frac{1}{N}\sum_{j=1}^{N}\hat\cn_{j,u^l,\delta}^l.  \]
We will show the existence of good codes for the reduced channels $\mathcal{Q}_{j,\delta,l}\circ\cn_{j,\delta,l}$ in the limit of large $l\in\mathbb N$. An application of Lemma \ref{gentle-operator-lemma-for-F_e} and Lemma \ref{lemma-typical-2} will then show that these codes are also good for the original compound channel.\\
Let $U^l$ be a random variable taking values in
$\mathfrak{U}(\gr^{\otimes l})$ which is distributed according to
the Haar measure. Application of Theorem \ref{convex-klesse}
yields
\begin{eqnarray}
\mathbb
E{F}_{c,e}(\pi_{\E_l},\hat\cn^l_{U^l,\delta})&\ge&\textrm{tr}(\hat\cn^l_\delta(\pi_{\gr}^{\otimes
l}))-2\sum_{j=1}^{N}\sqrt{k_ln_{j,\delta,l}}||\hat\cn_{j,\delta}^l(\pi_{\gr}^{\otimes
l})||_2,\label{eq:direct-finite-uninformed-users-1}
\end{eqnarray}
where $n_{j,\delta,l}$ stands for the number of Kraus operators of
the reduced operation $\cn_{j,\delta,l}$
($j\in\{1,\ldots,N\}$) and
\[ \hat\cn_{j,\delta}^l:=\mathcal Q_{j,\delta,l}\circ\cn_{j,\delta,l},  \]
\[\hat\cn^l_\delta:=\frac{1}{N}\sum_{j=1}^{N}\hat\cn_{j,\delta}^l.  \]
Notice that $\mathcal Q_{j,\delta,l}\circ\cn_{j,\delta,l}$ trivially has a Kraus representation containing exactly $n_{j,\delta,l}$ elements. We will use inequality (\ref{eq:direct-finite-uninformed-users-1}) in the proof
of the following theorem.
\begin{theorem}[Direct Part: Uninformed Users and $|\fri|<\infty$]\label{lemma-direct-finite-uninformed-users-1}\
Let $\fri=\{\cn_1,...,\cn_N\}\subset \mathcal{C}(\hr,\kr)$ be a
compound channel and $\pi_\gr$ the maximally mixed state
associated to a subspace $\gr\subset\hr$. Then
$$Q(\fri)\geq\min_{\cn_i\in\fri}I_c(\pi_\gr,\cn_i).$$
\end{theorem}
\emph{Proof}. We show that for every $\epsilon>0$ the number $\min_{\cn_i\in\fri}I_c(\pi_\gr,\cn_i)-\epsilon$ is an achievable rate for $\fri$.\\
1) If $\min_{\cn_i\in\fri}I_c(\pi_\gr,\cn_i)-\epsilon\leq0$, there is nothing to prove.\\
2) Let $\min_{\cn_i\in\fri}I_c(\pi_\gr,\cn_i)-\epsilon>0$.\\
Choose $\delta\in(0,1/2)$ and $l_0\in\mathbb N$ satisfying $\gamma(\delta)+\vphi(\delta)+h'(l_0)\leq\epsilon/2$ with functions $\gamma,\vphi,h'$ from Lemma \ref{lemma-typical-1} and \ref{lemma-typical-2}.\\
Now choose for every $l\in\mathbb N$ a subspace $\E_l\subset\gr^{\otimes l}$ such that
\[\dim \E_l=:k_l=\lfloor2^{l(\min_{\cn_i\in\fri}I_c(\pi_\gr,\cn_i)-\epsilon)}\rfloor.\]
By $S(\pi_\gr)\geq I_c(\pi_\gr,\cn_j)$ (see \cite{bkn}), this is always possible.\\
Obviously,
\[ \min_{\cn_i\in\fri}I_c(\pi_\gr,\cn_i)-\epsilon-o(l^0)\le  \frac{1}{l}\log k_l\leq\min_{\cn_i\in\fri}I_c(\pi_\gr,\cn_i)-\epsilon.\]
We will now give lower bounds on the terms in
(\ref{eq:direct-finite-uninformed-users-1}), thereby making use of
Lemma \ref{lemma-typical-1} and Lemma \ref{lemma-typical-2}:
\begin{eqnarray}\label{eq:direct-finite-uninformed-users-2}
\textrm{tr}(\hat\cn^l_\delta(\pi_{\gr}^{\otimes l})
)\geq1-2^{-l(c\delta^2-h(l))}-2^{-l(c'\delta^2-h'(l))}.
\end{eqnarray}
A more detailed calculation can be found in \cite{bbn-1} or
\cite{klesse}. Further, and additionally using the inequality
$||A+B||_2^2\geq||A||_2^2+||B||_2^2$ valid for non-negative
operators $A,B\in\mathcal{B}(\kr^{\otimes l})$ (see
\cite{klesse}), we get the inequality
\begin{eqnarray}\label{eq:direct-finite-uninformed-users-3}
||\hat\cn_{j,\delta}^l(\pi_{\gr}^{\otimes l})||_2^2&\leq&2^{-l(S(\cn_j(\pi_{\gr}))-\vphi(\delta))}.
\end{eqnarray}
From (\ref{eq:direct-finite-uninformed-users-1}),
(\ref{eq:direct-finite-uninformed-users-2}),
(\ref{eq:direct-finite-uninformed-users-3}) and our specific
choice of $k_l$ it follows that
\begin{eqnarray}
\mathbb EF_{c,e}(\pi_{\E_l},\hat\cn^l_{U^l,\delta})&\ge&1-2^{-l(c\delta^2-h(l))}-2^{-l(c'\delta^2-h'(l))}\nonumber\\
&&-2\sum_{j=1}^N\sqrt{2^{l(\frac{1}{l}\log k_l+\gamma(\delta)+\vphi(\delta)+h'(l)-I_c(\pi_{\gr},\cn_j)}}\nonumber\\
&\geq&1-2^{-l(c\delta^2-h(l))}-2^{-l(c'\delta^2-h'(l))}\nonumber\\
&& -2N\sqrt{2^{-l(\epsilon-\gamma(\delta)-\vphi(\delta)-h'(l))}}\nonumber.
\end{eqnarray}
Since $\epsilon-\gamma(\delta)-\vphi(\delta)-h'(l)\geq\eps/2$ for every $l\geq l_0$, this shows the
existence of at least one sequence of $(l,k_l)-$codes for
$\fri$ with uninformed users and
\[ \liminf_{l\rightarrow\infty}\frac{1}{l}\log k_l=\min_{\cn_i\in\fri}I_c(\pi_{\gr},\cn_i)-\epsilon \]
as well as (using that entanglement fidelity is affine in the channel), for every $l\in\mathbb N$,
\begin{equation} \min_{j\in\{1,\ldots,N\}}F_e(\pi_{\fr_l},\crr^l\circ\hat\cn_{j,\delta}^l\circ\mathcal W^l)\geq1-N\frac{1}{3}\epsilon_l \label{eq:direct-finite-uninformed-users-4}\end{equation}
where $w^l\in\mathfrak U(\gr^{\otimes l})\ \forall l\in\mathbb N$ and
\begin{equation}\label{eq:direct-finite-uninformed-users-eps}
\epsilon_l=3\cdot(2^{-l(c\delta^2-h(l))}+2^{-l(c'\delta^2-h'(l))}+2N\sqrt{2^{-l(\epsilon-\gamma(\delta)-\vphi(\delta)-h'(l))}}).
\end{equation}
Note that $\lim_{l\to\infty}\epsilon_l=0$ exponentially fast, as can be seen from our choice of $\delta$ and $l_0$.
For every $j\in\{1,\ldots,N\}$ and $l\in\mathbb N$ we thus have,
by property $3.$ of Lemma \ref{lemma-typical-2}, construction of
$\hat\cn_{j,w^j,\delta}^l$, and equation
(\ref{eq:direct-finite-uninformed-users-4}),
\begin{eqnarray*}
F_e(\pi_{\fr_l},\crr^l\circ\mathcal Q_{j,\delta,l}\circ\cn_{j}^{\otimes l}\circ\mathcal W^l)&\geq&F_e(\pi_{\fr_l},\crr^l\circ\mathcal Q_{j,\delta,l}\circ\cn_{j,\delta,l}\circ\mathcal W^l)\\
&=&F_e(\pi_{\fr_l},\crr^l\circ\hat\cn_{j,w^j,\delta}^l)\\
&\geq&1-N\frac{1}{3}\epsilon_l.
\end{eqnarray*}
By the first two parts of Lemma \ref{gentle-operator-lemma-for-F_e}, this immediately
implies
\begin{equation}\label{eq:direct-finite-uninformed-users-5}
\min_{\cn_j\in\fri}F_e(\pi_{\fr_l},\crr^l\circ\cn_{j}^{\otimes l}\circ\mathcal W^l)\geq1-N\epsilon_l\ \ \ \forall l\in\mathbb N.
\end{equation}
Since $\epsilon>0$ was arbitrary, we have shown that
$\min_{\cn_i\in\fri}I_c(\pi_{\gr},\cn_i)$ is an achievable rate.
\begin{flushright}$\Box$\end{flushright}
\subsection{\label{subsec:The Informed Encoder}The Informed Encoder}
In this subsection we shall prove the following Theorem:
\begin{theorem}[Direct Part: Informed Encoder and $| \fri |<\infty $]\label{direct-finite-informed-encoder}
For every finite compound channel
$\fri=\{\cn_1,\ldots,\cn_N\}\subset\mathcal C(\hr,\kr)$ and any
set $\{\pi_{\gr_1},\ldots,\pi_{\gr_N}\}$ of maximally mixed states
on subspaces $\{\gr_1,\ldots,\gr_N\}$ with $\gr_i\subset\hr$ for all $ i\in\{1,\ldots,N\}$ we have
$$Q_{IE}(\fri)\geq\min_{\cn_i\in\fri}I_c(\pi_{\gr_i},\cn_i).$$
\end{theorem}
\emph{Proof.} Let a compound channel be given by a finite set $\fri:=\{\cn_1,\ldots ,\cn_{N} \}\subset \mathcal{C}(\hr,\kr)$ and let $\gr_1,\ldots,\gr_N$ be arbitrary subspaces of $\hr$. We will prove that for every $\epsilon>0$ the value
$$R(\epsilon):=\min_{1\leq i\leq N}I_c(\pi_{\gr_i},\cn_i)-\epsilon$$
is achievable. If $R(\epsilon)\leq0$, there is nothing to prove. Hence we assume $R(\epsilon)>0$. For every $l\in\mathbb N$ and all $i\in\{1,\ldots,N\}$ we choose the following. First, a subspace $\E_l\subset\hr^{\otimes l}$ of dimension $k_l:=\dim\E_l$ that satisfies $k_l\leq\dim\gr_i^{\otimes l}$. Second, a set $\{v_1^l,\ldots,v_N^l\}$ of unitary operators  with the property $v_i^l\mathcal{E}_l\subset\gr^{\otimes l}_i$. Again, the maximally mixed states associated to the above mentioned subspaces are denoted by $\pi_{\mathcal{E}_l}$ on $\mathcal{E}_l$ and  $\pi_{\gr_i}$ on $\gr_i$.\\
For $j\in\{1,\ldots,N\}$, $\delta\in(0,1/2)$, $l\in\nn$ and states $\cn_j(\pi_{\gr_j})$ let $q_{j,\delta,l}\in \mathcal{B}(\kr)^{\otimes l}$ be the frequency-typical projection of $\cn_j(\pi_{\gr_j})$ and $\cn_{j,\delta,l}$ be the reduced operation associated with $\cn_j$ and $\pi_{\gr_j}$ as considered in section \ref{subsec:Typical Projections and Kraus Operators}.\\
Let, for the moment, $l\in\mathbb N$ be fixed. We define a new set of channels that is more adapted to our problem than the original one. We set, for an arbitrary set $\{u_1^l,\ldots,u_N^l\}$ of unitary operators on $\hr^{\otimes l}$
\begin{eqnarray*}
\tilde\cn^l_{j,\delta}:=\mathcal{Q}_{j,\delta,l}\circ\cn_{j,\delta,l},
\end{eqnarray*}
\begin{eqnarray*}
\hat\cn^l_{j,u_j^l,\delta}:=\tilde\cn^l_{j,\delta}\circ\U_j^l\circ\V_j^l
\end{eqnarray*}
and, accordingly,
\[\hat\cn^l_{u_1^l,\ldots,u_N^l,\delta}:=\frac{1}{N}\sum_{j=1}^{N}\hat\cn^l_{j,u_j^l,\delta}.  \]
we will first show the existence of good unitary encodings and recovery operation for $\{\tilde\cn^l_{1,\delta},\ldots,\tilde\cn^l_{N,\delta}\}$. Like in the previous subsection, application of Lemma \ref{gentle-operator-lemma-for-F_e} will enable us to show the existence of reliable encodings and recovery operation for the original compound channel $\fri$. \\
Let $U_1^l,\ldots,U_N^l$ be independent random variables such that
each $U_i^l$ takes on values in $\mathfrak U(\gr_i^{\otimes l})$
and is distributed according to the Haar measure on $\mathfrak
U(\gr_i^{\otimes l})$ $(i\in\{1,\ldots,N\}$). By Theorem
\ref{uie-theorem} we get the lower bound
\begin{eqnarray}
\mathbb EF_{c,e}(\pi_{\E_l},\hat\cn^l_{U_1^l,\ldots,U_N^l,\delta})\ge
\sum_{j=1}^N[\frac{1}{N}\textrm{tr}(\tilde\cn^l_{j,\delta}(\pi_{\gr_j^{\otimes
l}}))- 2\sqrt{k_l n_{j,\delta,l} }||\tilde\cn^l_{j,\delta}
(\pi_{\gr_j^{\otimes l}})||_2],
\end{eqnarray}
where $n_{j,\delta,l}$ denotes the number of Kraus operators in
the operations $\tilde\cn_{j,\delta,l}$ ($j\in\{1,\ldots,N\}$). By
Lemmas \ref{lemma-typical-1},\ref{lemma-typical-2} for every
$j\in\{1,\ldots,N\}$ the corresponding term in the above sum can
be bounded from below through
\begin{eqnarray*}
\frac{1}{N}\textrm{tr}(\tilde\cn^l_{j,\delta}(\pi_{\gr_j^{\otimes
l}}))\geq\frac{1}{N}(1-2^{-l(c\delta^2-h(l))}-2^{-l(c'\delta^2-h'(l))})
\end{eqnarray*}
and
\begin{eqnarray*}
-2\sqrt{k_l n_{j,\delta,l} }||\tilde\cn^l_{j,\delta} (\pi_{\gr_j^{\otimes
l}})||_2\geq-2\sqrt{k_l\cdot2^{l(-\min_{1\leq j\leq N}I_c(\pi_{\gr_j},\cn_j)+\gamma(\delta)+\vphi(\delta)+h'(l))}}.
\end{eqnarray*}
Set $k_l:=\lfloor2^{lR(\epsilon)}\rfloor$. Obviously, for any $j\in\{1,\ldots,N\}$,
$$k_l\cdot2^{l(-\min_{1\leq j\leq N}I_c(\pi_{\gr_j},\cn_j))}\leq2^{-l\epsilon}.$$
This implies
\begin{eqnarray*}
\mathbb
EF_{c,e}(\pi_{\E_l},\hat\cn^l_{U_1^l,\ldots,U_N^l,\delta})&\ge&1-2^{-l(c\delta^2-h(l))}-2^{-l(c'\delta^2-h'(l))}\\
&&-2N\sqrt{2^{l(-\epsilon+\gamma(\delta)+\vphi(\delta)+h'(l))}}.
\end{eqnarray*}
Now choosing both the approximation parameter $\delta$ and an integer $l_0\in\mathbb N$ such that $-\epsilon+\gamma(\delta)+\vphi(\delta)+h'(l)<-\frac{1}{2}\epsilon$ holds for every $l\geq l_0$ and setting $$\epsilon_l:=2^{-l(c\delta^2-h(l))}+2^{-l(c'\delta^2-h'(l))}+2N\sqrt{2^{l(-\epsilon+\gamma(\delta)+\vphi(\delta)+h'(l))}}$$
we see that
\begin{eqnarray*}
\mathbb
E F_{c,e}(\pi_{\E_l},\hat\cn^l_{U_1^l,\ldots,U_N^l,\delta})  &\ge&1-\epsilon_l,
\end{eqnarray*}
where again $\epsilon_l\searrow0$ and
our choice of $\delta$ and $l_0$ again shows that the speed of convergence is exponentially fast.
Thus, there exist unitary operators
$w_1^l,\ldots,w_N^l\subset\mathfrak U(\hr^{\otimes l})$ and a
recovery operation $\crr^l$ such that, passing to the individual
channels, we have for every $j\in\{1,\ldots,N\}$
\[F_{e}(\pi_{\E_l},\crr^l\circ\mathcal
Q_{j,\delta,l}\circ\cn_{j,\delta,l}\circ\mathcal
W_j^l)\ge 1-N\epsilon_l.\]
By property $3.$ of Lemma \ref{lemma-typical-2} and Lemma \ref{gentle-operator-lemma-for-F_e}, we immediately see that
\begin{eqnarray*}
F_{e}(\pi_{\E_l},\crr^l\circ\cn_{j}^{\otimes l}\circ\mathcal
W_j^l)&\ge&1-3N\epsilon_l\ \ \ \forall
j\in\{1,\ldots,N\}
\end{eqnarray*}
is valid as well. We finally get the desired result: For every set
$\{\pi_{\gr_1},\ldots,\pi_{\gr_N}\}$ of maximally mixed states on
subspaces $\gr_1,\ldots,\gr_N\subset\hr$ and every $\epsilon>0$
there exists a sequence of $(l,k_l)$ codes for $\fri$ with
informed encoder with the properties
\begin{enumerate}
\item $\liminf_{l\rightarrow\infty}\frac{1}{l}\log
k_l=\min_{\cn_j\in\fri}I_c(\pi_{\gr_j},\cn_j)-\epsilon$,
\item
$\min_{\cn_j\in\fri}F_e(\pi_{\E_l},\crr^l\circ\cn_j^{\otimes
l}\circ\mathcal W^l_j)\geq1-3N\epsilon_l$.
\end{enumerate}
Since $\epsilon>0$ was arbitrary and
$\epsilon_l\searrow0$, we are done.
\begin{flushright}$\Box$\end{flushright}

\section{\label{sec:discretization}Finite Approximations in the Set of Quantum Channels}
Our goal in this section is to discretize a given set of channels $\fri\in\mathcal{C}(\hr,\kr)$ in such a way that the results derived so far for finite sets can be employed to derive general versions of coding theorems for compound channels.\\
The first concept we will need is that of a $\tau$-net in the set $\mathcal{C}(\hr,\kr)$ and we will give an upper bound on the cardinality of the best $\tau$-net in that set. Best $\tau$-nets characterize the degree of compactness of $\mathcal{C}(\hr,\kr)$.\\
A $\tau$-net in $\mathcal{C}(\hr,\kr)$ is a finite set $\{ \cn_i \}_{i=1}^{N}$ with the property that for each $\cn\in\mathcal{C}(\hr,\kr)$ there is at least one $i\in\{1,\ldots,N  \}$ with $||\cn-\cn_i||_{\lozenge}<\tau$. Existence of $\tau$-nets in $\mathcal{C}(\hr,\kr)$ is guaranteed by the compactness of $\mathcal{C}(\hr,\kr)$. The next lemma contains a crude upper bound on the cardinality of minimal $\tau$-nets.
\begin{lemma}\label{lemma-tau-nets}
For any $\tau\in (0,1]$ there is a $\tau-$net $\{\cn_i  \}_{i=1}^{N}$ in $\mathcal{C}(\hr,\kr)$ with $N\le (\frac{3}{\tau})^{2(d\cdot d')^2}$, where $d=\dim \hr$ and $d'=\dim \kr$.
\end{lemma}
\emph{Proof.} The assertion of the lemma follows from the standard volume argument (cf. Lemma 2.6 in \cite{milman}). The details can be found in our previous paper \cite{bbn-1}.
\begin{flushright}$\Box$\end{flushright}
Let $\fri\subseteq \mathcal{C}(\hr,\kr)$ be an arbitrary set. Starting from a $\tau/2-$net $\mathfrak{N}:=\{ \cn_i \}_{i=1}^{N}$ with $N\le (\frac{6}{\tau})^{2(d\cdot d')^2} $ as in Lemma \ref{lemma-tau-nets} we can build a $\tau/2-$net $\fri'_{\tau}$ that is adapted to the set $\fri$ given by
\begin{equation}\label{adapted-net-prel}
  \fri'_{\tau}:=\left\{\cn_i\in \mathfrak{N}: \exists \cn\in \fri \textrm{ with }||\cn-\cn_i||_{\lozenge}<\tau/2  \right\},
\end{equation}
i.e. we select only those members of the $\tau/2$-net that are contained in the $\tau/2$-neighborhood of $\fri$. Let $\mathcal{T}\in\mathcal{C}(\hr,\kr) $ be the useless channel given by $\mathcal{T}(\rho):=\frac{1}{\dim \kr}\idn_{\kr}$, $\rho\in \mathcal{S}(\hr)$, and consider
\begin{equation}\label{adapted-net}
  \fri_{\tau}:=\left\{(1-\frac{\tau}{2})\cn+\frac{\tau}{2} \mathcal{T}:\cn\in\fri'_{\tau}  \right\},
\end{equation}
where $\fri'_{\tau}$ is defined in (\ref{adapted-net-prel}).
For $\fri\subseteq \mathcal{C}(\hr,\kr) $ we set
\[I_{c}(\rho, \fri ):=\inf_{\cn\in \fri}I_{c}(\rho,\cn),  \]
for $\rho\in \cs (\hr)$.
We list a few more or less obvious results in the following lemma that will be needed in the following.
\begin{lemma}\label{lemma-approx-properties}
Let $\fri\subseteq \mathcal{C}(\hr,\kr) $. For each positive $\tau\le \frac{1}{e}$ let $\fri_{\tau}$ be the finite set of channels defined in (\ref{adapted-net}).
\begin{enumerate}
\item $|\fri_{\tau}|\le (\frac{6}{\tau})^{2(d\cdot d')^2} $ with $d=\dim \hr$ and $d'=\dim \kr$.
\item For $\cn\in \fri$ there is $\cn_i\in \fri_{\tau}$ with
\begin{equation}\label{eq:approx-prop-1}
||\cn^{\otimes l}-\cn_i^{\otimes l}||_{\lozenge}< l\tau.
\end{equation}
Consequently, for $\cn$, $\cn_i$, and any CPTP maps $\mathcal{P}: \mathcal{B}(\fr)\to \mathcal{B}(\hr)^{\otimes l}$  and $\crr: \mathcal{B}(\kr)^{\otimes l}\to\mathcal{B}(\mathcal{F}')$ the relation
\begin{equation}\label{eq:approx-prop-2}
 |F_e(\rho,\crr\circ \cn^{\otimes l}\circ \mathcal{P})- F_e(\rho,\crr\circ \cn_i^{\otimes l}\circ\mathcal{P})|<l\tau
\end{equation}
holds for all $\rho\in\cs(\hr^{\otimes l})$ and $l\in \nn$.
\item For all $\rho\in\cs (\hr)$ we have
  \begin{equation}\label{eq:approx-prop-3}
    |I_{c}(\rho,\fri)-I_{c}(\rho,\fri_{\tau})|\le \tau+3\tau\log \frac{d}{\tau}.
  \end{equation}

\end{enumerate}
\end{lemma}
\emph{Proof.} The proofs of the assertions claimed here are either identical to those given in \cite{bbn-1} or can be obtained by trivial modifications thereof.
\begin{flushright}$\Box$\end{flushright}
\section{\label{sec:Direct Parts of the Coding Theorems for General Quantum Compound Channels}Direct Parts of the Coding Theorems for General Quantum Compound Channels}

\subsection{\label{subsec:The Case of Informed Decoder and Uninformed Users}The Case of Informed Decoder and Uninformed Users}
The main step towards the direct part of the coding theorem for quantum compound channels with uninformed users is the following theorem.
\begin{lemma}\label{direct-coding-general-uu}
Let $\fri\in \mathcal{C}(\hr,\kr)$ be an arbitrary compound channel and let $\pi_{\gr}$ be the maximally mixed state associated with a subspace $\gr\subset \hr$. Then
\[Q(\fri)\ge \inf_{\cn\in \fri}I_c(\pi_{\gr},\cn).  \]
\end{lemma}
\emph{Proof.} We consider two subspaces $\E_l, \gr^{\otimes l}$ of $\hr^{\otimes l}$ with $\E_l\subset\gr^{\otimes l}\subset \hr^{\otimes l}$. Let $k_l:=\dim \E_l$ and we denote as before the associated maximally mixed states on $\E_l$ and $\gr$ by $\pi_{\E_l}$ and  $\pi_{\gr}$.\\
If $\inf_{\cn\in \fri}I_c(\pi_{\gr},\cn)\le 0 $ there is nothing to prove. Therefore we will suppose in the following that
\[ \inf_{\cn\in \fri}I_c(\pi_{\gr},\cn)>0 \]
holds. We will show that for each $\eps\in (0, \inf_{\cn\in \fri}I_c(\pi_{\gr},\cn) )$ the number
\[\inf_{\cn\in \fri}I_c(\pi_{\gr},\cn)-\eps  \]
is an achievable rate.\\
For each $l\in\nn$ let us choose some $\tau_l>0$ with $\tau_l\le \frac{1}{e}$, $\lim_{l\to\infty}l\tau_l=0$, and such that $N_{\tau_l}$ grows sub-exponentially with $l$. E.g. we may choose $\tau_l:=\min\{1/e, 1/l^2\}  $. We consider, for each $l\in\nn$, the finite set of channels $\fri_{\tau_l}:=\{\cn_1,\ldots, \cn_{N_{\tau_l}}\}$ associated to $\fri$ given in (\ref{adapted-net}) with the properties listed in Lemma \ref{lemma-approx-properties}.
We can conclude from the proof of Theorem \ref{lemma-direct-finite-uninformed-users-1} that for each $l\in \nn$ there is a subspace $\fr_l\subset \gr^{\otimes l}$ of dimension
\begin{equation}\label{eq:direct-coding-uu-1}
k_l= \lfloor  2^{l(\min_{i\in \{1,\ldots, N_{\tau}  \}}I_c(\pi_{\gr},\cn_i)-\frac{\eps}{2}  )}  \rfloor,
\end{equation}
 a recovery operation $\mathcal{R}$, and a unitary encoder $\mathcal{W}^l$ such that
\begin{eqnarray}\label{eq:direct-coding-uu-2}
\min_{i\in\{1,\ldots,N_{\tau_l}  \}}F_e(\pi_{\fr_l}, \mathcal{R}\circ \cn_i^{\otimes l}\circ \mathcal{W}^l )&\ge& 1-N_{\tau_{l}}\epsilon_{l}
\end{eqnarray}
where $\epsilon_l$ is defined in (\ref{eq:direct-finite-uninformed-users-eps}) (with the approximation parameter $\eps$ replaced by $\eps/2$), and we have chosen $l,l_0\in\mathbb N$ with $l\geq l_0$ large enough and $\delta>0$ small enough to ensure that both
\[ \min_{i\in \{1,\ldots, N_{\tau_l}  \}}I_c(\pi_{\gr},\cn_i)-\frac{\eps}{2}>0,  \]
and
\[ \frac{\eps}{2}-\gamma(\delta)-\vphi(\delta)-h'(l_0)>\eps/4>0. \]
By our construction of $\fri_{\tau_l}$ we can find to each $\cn\in \fri$ at least one $\cn_i\in \fri_{\tau_l}$ with
\begin{equation}\label{eq:direct-coding-uu-3}
 |F_{e}(\pi_{\fr_l},\mathcal{R}\circ \cn_i^{\otimes l}\circ \mathcal{W}^l )- F_{e}(\pi_{\fr_l},\mathcal{R}\circ \cn^{\otimes l}\circ \mathcal{W}^l )     | \le l\cdot \tau_l
\end{equation}
according to Lemma \ref{lemma-approx-properties}. Moreover, by the last claim of Lemma \ref{lemma-approx-properties} we obtain the following estimate on the dimension $k_l$ of the subspace $\fr_l$:
\begin{equation}\label{eq:direct-coding-uu-4}
k_l \ge \lfloor  2^{l (\inf_{\cn\in \fri}I_c(\pi_{\gr},\cn )-\frac{\eps}{2}-\tau_l -2\tau_l\log\frac{d}{\tau_l}}   \rfloor.
\end{equation}
The inequalities (\ref{eq:direct-coding-uu-2}) and (\ref{eq:direct-coding-uu-3}) show that
\[\inf_{\cn\in\fri} F_e(\pi_{\fr_l}, \mathcal{R}\circ \cn^{\otimes l}\circ \mathcal{W}^l )\ge 1-N_{\tau_l}\epsilon_l-l\tau_l, \]
which in turn with (\ref{eq:direct-coding-uu-4}) shows that $\inf_{\cn\in \fri}I_c(\pi_{\gr},\cn ) $ is an achievable rate.
\begin{flushright}$\Box$\end{flushright}
In order to pass from the maximally mixed state $\pi_{\gr}$ to an arbitrary one we have to employ the compound generalization of Bennett, Shor, Smolin, and Thapliyal Lemma (BSST Lemma for short) from \cite{bsst} and \cite{holevo-ent-ass-cap}. For the proof of this generalized BSST Lemma we refer to \cite{bbn-1}.
\begin{lemma}[Compound BSST Lemma]\label{compound-bsst-lemma}
Let $\fri\subset\mathcal{C}(\hr,\kr) $ be an arbitrary set of channels. For any $\rho\in \mathcal{S}(\hr)$ let $q_{\delta,l}\in\mathcal{B}(\hr^{\otimes l})$ be the frequency-typical projection of $\rho$ and set
\[ \pi_{\delta,l}:=\frac{q_{\delta,l}}{\textrm{tr}(q_{\delta,l})}\in\mathcal{S}(\hr^{\otimes l}). \]
Then there is a positive sequence $(\delta_l)_{l\in\nn}$ satisfying $\lim_{l\to\infty}\delta_l=0$ with
\[ \lim_{l\to\infty}\frac{1}{l}\inf_{\cn\in \fri}I_c(\pi_{\delta_l,l},\cn^{\otimes l} )=\inf_{\cn\in\fri}I_c(\rho,\cn).  \]

\end{lemma}
With these preparations it is easy now to finish the proof of the direct part of the coding theorem for the quantum compound channel with uninformed users.\\
First notice that for each $k\in\nn$
\begin{equation}\label{eq:final-steps-uu-1}
Q(\fri^{\otimes k})=k Q(\fri)
\end{equation}
holds. For any fixed $\rho\in\mathcal{S}(\hr^{\otimes m})$ let $q_{\delta,l}\in \mathcal{B}(\hr^{\otimes ml})$ be the frequency-typical projection of $\rho$ and set $\pi_{\delta,l}=\frac{q_{\delta,l}}{\textrm{tr}(q_{\delta,l})}$. Lemma \ref{direct-coding-general-uu} implies that for any $\delta \in (0,1/2) $ we have
\begin{equation}\label{eq:final-steps-uu-2}
Q(\fri^{\otimes ml})\ge I_c(\pi_{\delta,l},\fri^{\otimes ml}),
\end{equation}
for all $m,l\in\nn$. Utilizing (\ref{eq:final-steps-uu-1}), (\ref{eq:final-steps-uu-2}) and Lemma \ref{compound-bsst-lemma} we arrive at
\begin{eqnarray}\label{eq:final-steps-uu-3}
 Q(\fri)&=& \frac{1}{m}\lim_{l\to\infty}\frac{1}{l}Q(\fri^{\otimes ml})\nonumber\\
        &\ge& \frac{1}{m}  \lim_{l\to\infty}\frac{1}{l}\inf_{\cn\in \fri}I_c(\pi_{\delta_l,l},(\cn^{\otimes m})^{\otimes l} )\nonumber\\
        &=&  \frac{1}{m}I_c(\rho, \fri^{\otimes m}).
\end{eqnarray}
From (\ref{eq:final-steps-uu-3}) and since $Q_{ID}(\fri)\geq Q(\fri)$ trivially holds we get without further ado the direct part of the coding theorem.
\begin{theorem}[Direct Part: Informed Decoder and Uninformed Users]\label{final-direct-part-uu}
Let $\fri\subset\mathcal{C}(\hr,\kr)$ be an arbitrary set. Then
\begin{equation}\label{eq:final-direct-part-uu}
Q_{ID}(\fri)\geq Q(\fri)\ge \lim_{l\to\infty}\frac{1}{l} \max_{\rho\in\mathcal{S}(\hr^{\otimes l})}\inf_{\cn\in\fri}I_c(\rho,\cn^{\otimes l}).
\end{equation}
\end{theorem}
\begin{remark}
It is quite easy to see that the limit in (\ref{eq:final-direct-part-uu}) exists. Indeed it holds that
\begin{eqnarray*}
  \max_{\rho\in\mathcal{S}(\hr^{\otimes l+k})}\inf_{\cn\in\fri}I_c(\rho,\cn^{\otimes l+k})&\ge& \max_{\rho\in\mathcal{S}(\hr^{\otimes l})}\inf_{\cn\in\fri}I_c(\rho,\cn^{\otimes l})\\
&&+ \max_{\rho\in\mathcal{S}(\hr^{\otimes k})}\inf_{\cn\in\fri}I_c(\rho,\cn^{\otimes k})
\end{eqnarray*}
which implies the existence of the limit via standard arguments.
\end{remark}
\subsection{\label{subsec:The Informed Encoder}The Informed Encoder}
The main result of this section will rely on an appropriate variant of the BSST Lemma. To this end we first recall Holevo's version of that result. For $\delta>0$, $l\in\nn$, and $\rho\in\cs (\hr)$ let $q_{\delta,l}\in\mathcal{B}(\hr^{\otimes l})$ denote the frequency typical projection of $\rho^{\otimes l}$. Set
\begin{equation}\label{eq:def-flat-state}
\pi_{\delta,l}=\pi_{\delta,l}(\rho):=\frac{q_{\delta,l} }{\textrm{tr}(q_{\delta,l})}.  
\end{equation}
Moreover, let
\[\lambda_{\min}(\rho):=\min\{\lambda\in \sigma(\rho): \lambda>0  \},   \]
where $\sigma(\rho)$ stands for the spectrum of the density operator $\rho$. 

\begin{lemma}[BSST Lemma \cite{bsst}, \cite{holevo-ent-ass-cap}]\label{holevo-bsst-lemma}
 For any $\delta \in (0,\frac{1}{2\dim \hr})$, any $\cn\in\mathcal{C}(\hr,\kr)$, and every $\rho\in\cs(\hr)$ with associated state $\pi_{\delta,l}=\pi_{\delta,l}(\rho)\in\cs (\hr^{\otimes l})$ we have
  \begin{equation}\label{eq:holevo-bsst-lemma}
    \left|\frac{1}{l}S(\cn^{\otimes l}(\pi_{\delta,l}))-S(\cn(\rho))    \right| \le \theta_l(\delta,\lambda_{\min}(\rho),\lambda_{\min}(\cn(\rho)))
  \end{equation}
where
\begin{eqnarray}\label{eq:holevo-bsst-lemma-rhs}
\theta_l(\delta,\lambda_{\min}(\rho),\lambda_{\min}(\cn(\rho)))  &=& \frac{\dim \hr}{l}\log (l+1)-\dim \hr\cdot \delta\log\delta\nonumber\\
&&-\dim \hr\cdot\delta\cdot ( \log \lambda_{\min}(\rho)
+ \log \lambda_{\min}(\cn(\rho))).
\end{eqnarray}
\end{lemma}
Before we present our extended version of BSST Lemma we introduce some notation. For $t\in (0,\frac{1}{e})$ and any set $\fri\subset \mathcal{C}(\hr,\kr)$ let us define
\begin{equation}\label{eq:def-I-t}
\fri^{(t)}:=\{\cn^{(t)}=(1-t)\cn +t \mathcal{T}_{\kr}: \cn\in\fri  \}=(1-t)\fri+ t\mathcal{T}_{\kr},  
\end{equation}
where $\mathcal{T}\in \mathcal{C}(\hr,\kr) $ is given by $\mathcal{T}_{\kr}(x):=\frac{\tr (x)}{\dim \kr}\idn_{\kr}$.\\
On the other hand, to each $\cn\in \fri\subset \mathcal{C}(\hr,\kr)$ we can associate a complementary channel $\cn_c\in\mathcal{C}(\hr,\hr_e)$ where we assume w.l.o.g. that $\hr_e=\cc^{\dim \hr \cdot \dim \kr} $. Let $\fri'\subset \mathcal{C}(\hr,\hr_e)$ denote the set of channels complementary to $\fri$ and set
\begin{equation}\label{eq:def-I-ct}
  \fri'^{(t)}:= (\fri')^{(t)}=\{\cn_c^{(t)}=(1-t)\cn_{c}+t \mathcal{T}_{\hr_e}:\cn_c\in\fri'  \}=(1-t)\fri' +t \mathcal{T}_{\hr_e},
\end{equation}
where $\mathcal{T}_{\hr_e}\in\mathcal{C}(\hr,\hr_e)$ is the defined in a similar way as $\mathcal{T}_{\kr} $. Finally, for $\cn\in\fri$ let
\[\rho_{\cn}:=\arg\max_{\rho\in\cs (\hr)}I_c(\rho,\cn),  \]
and for $t\in(0,\frac{1}{e})$, $\delta>0$, and $l\in\nn$ define
\begin{equation}\label{eq:def-pi-cn}
  \pi_{\delta,l,\cn}^{(t)}:=\pi_{\delta,l}\left( \rho_{\cn}^{(t)}\right),
\end{equation}
where we have used the notation from (\ref{eq:def-flat-state}) and
\begin{equation}\label{eq:def-perturbed-state}
  \rho_{\cn}^{(t)}:=(1-t)\rho_{\cn} + \frac{t}{\dim \hr}\idn_{\hr}.
\end{equation}
\begin{lemma}[Uniform BSST-Lemma]\label{compound-bsst-ie-lemma}
1. Let $l\in\nn$, $t\in (0,\frac{1}{l\cdot e})$, and $\delta\in (0,\frac{1}{2\dim \hr})$. Then with the notation introduced in the preceding paragraph we have
\[ \left|\frac{1}{l}\inf_{\cn\in\fri}I_c(\pi_{\delta,l,\cn}^{(t)},\cn^{\otimes l} )-\inf_{\cn\in\fri}\max_{\rho\in\cs(\hr)}I_c(\rho,\cn)      \right|\le \Delta_l(\delta,t),   \]
with 
\begin{eqnarray*}
  \Delta_l(\delta,t)&=& 2\theta_l\left(\delta,\frac{t}{\dim\hr},\frac{t}{\dim \kr}\right)+ 2\theta_l\left(\delta,\frac{t}{\dim\hr},\frac{t}{\dim \hr_e}\right)\\
&&- 4t \log\frac{t}{\dim\kr\cdot \dim\hr_e} -2 lt\log \frac{l t}{\dim\kr\cdot \hr_e},\\
\end{eqnarray*}
where $\theta_l\left(\delta,\frac{t}{\dim\hr},\frac{t}{\dim \kr}\right) $ and $\theta_l\left(\delta,\frac{t}{\dim\hr},\frac{t}{\dim \hr_e}\right) $ are from Lemma \ref{holevo-bsst-lemma}.\\
2. Consequently, choosing suitable positive sequences $(\delta_l)_{l\in\nn}$, $(t_l)_{l\in\nn}$ with
\begin{enumerate}
\item $\lim_{l\to\infty}\delta_l=0=\lim_{l\to\infty}lt_l$, and
\item $\lim_{l\to\infty}\delta_l\log t_l=0$
\end{enumerate}
 we see that for $\nu_l:=\Delta_l(\delta_l,t_l) $
\begin{equation}\label{eq:compound-bsst-ie-lemma}
 \left|\frac{1}{l}\inf_{\cn\in\fri}I_c(\pi_{\delta_{l},l,\cn}^{(t_{l})},\cn^{\otimes l} )-\inf_{\cn\in\fri}\max_{\rho\in\cs(\hr)}I_c(\rho,\cn)      \right|\le \nu_l  
\end{equation}
holds with $\lim_{l\to\infty}\nu_l=0$.
\end{lemma}
\emph{Proof.} Our proof strategy is to reduce the claim to the BSST Lemma \ref{holevo-bsst-lemma}. Let $t>0$ be small enough to ensure that $l\cdot t\in (0,\frac{1}{e})$ and let $\delta\in (0,\frac{1}{2\dim \hr})$ be given. From (\ref{eq:def-I-t}) and (\ref{eq:def-I-ct}) we obtain that
\begin{equation}\label{eq:compound-bsst-ie-lemma-1}
  \lambda_{\min}(\cn^{(t)}(\rho))\ge \frac{t}{\dim \kr}, \qquad \lambda_{\min}(\cn_c^{(t)}(\rho) )\ge \frac{t}{\dim \hr_e}\qquad \forall\ \rho\in\cs(\hr),
\end{equation}
and (\ref{eq:def-perturbed-state}) yields that
\begin{equation}\label{eq:compound-bsst-ie-lemma-2}
  \lambda_{\min}(\rho_{\cn}^{(t)})\ge \frac{t}{\dim \hr}
\end{equation}
for all $\cn\in\fri$. The bounds (\ref{eq:compound-bsst-ie-lemma-1}) and (\ref{eq:compound-bsst-ie-lemma-2}) along with Lemma \ref{holevo-bsst-lemma} show that
\begin{equation}\label{eq:compound-bsst-ie-lemma-3}
  \left|\frac{1}{l}S((\cn^{(t)})^{\otimes l}(\pi_{\delta,l,\cn}^{(t)} ) ) -S(\cn^{(t)}(\rho_{\cn}^{(t)}))   \right|\le \theta_l\left(\delta,\frac{t}{\dim\hr},\frac{t}{\dim \kr}\right),
\end{equation}
and
\begin{equation}\label{eq:compound-bsst-ie-lemma-4}
  \left|\frac{1}{l}S((\cn_c^{(t)})^{\otimes l}(\pi_{\delta,l,\cn}^{(t)} ) ) -S(\cn_c^{(t)}(\rho_{\cn}^{(t)}))   \right|\le \theta_l\left(\delta,\frac{t}{\dim\hr},\frac{t}{\dim \hr_e}\right).
\end{equation}
On the other hand, by definition we have 
\begin{equation}\label{eq:compound-bsst-ie-lemma-5}
  ||\cn^{(t)}-\cn||_{\lozenge}\le t,\qquad ||(\cn^{(t)})^{\otimes l}-\cn^{\otimes l}    ||_{\lozenge}\le l\cdot t,
\end{equation}
and similarly
\begin{equation}\label{eq:compound-bsst-ie-lemma-6}
  ||\cn_c^{(t)}-\cn_c||_{\lozenge}\le t,\qquad ||(\cn_c^{(t)})^{\otimes l}-\cn_c^{\otimes l}    ||_{\lozenge}\le l\cdot t,
\end{equation}
for all $\cn\in\fri$. Since $l\cdot t\in (0,\frac{1}{e}) $ we obtain from this by Fannes inequality
\begin{equation}\label{eq:compound-bsst-ie-lemma-7}
  |S(\cn^{(t)}(\rho_{\cn}^{(t)}) )- S(\cn(\rho_{\cn}^{(t)}) )|\le - t \log\frac{t}{\dim\kr},
\end{equation}
\begin{equation}\label{eq:compound-bsst-ie-lemma-7a}
 |S(\cn_c^{(t)}(\rho_{\cn}^{(t)}) )- S(\cn_c(\rho_{\cn}^{(t)}) )|\le - t \log\frac{t}{\dim\hr_e}  
\end{equation}

and
\begin{equation}\label{eq:compound-bsst-ie-lemma-8}
 \left| \frac{1}{l}S((\cn^{(t)})^{\otimes l}(\pi_{\delta,l,\cn}^{(t)} ) )-\frac{1}{l}S(\cn^{\otimes l}(\pi_{\delta,l,\cn}^{(t)} ) ) \right|\le -l\cdot t\log \frac{l\cdot t}{\dim\kr},
\end{equation}
as well as
\begin{equation}\label{eq:compound-bsst-ie-lemma-9}
 \left| \frac{1}{l}S((\cn_c^{(t)})^{\otimes l}(\pi_{\delta,l,\cn}^{(t)} ) )-\frac{1}{l}S(\cn_c^{\otimes l}(\pi_{\delta,l,\cn}^{(t)} ) ) \right|\le -l\cdot t\log \frac{l\cdot t}{\dim\hr_e},
\end{equation}
for all $\cn\in\fri$. Since
\[I_c(\rho_{\cn}^{(t)},\cn )=S(\cn (\rho_{\cn}^{(t)}) )-S(\cn_c(\rho_{\cn}^{(t)}))  \]
and
\[I_c(\pi_{\delta,l,\cn}^{(t)},\cn^{\otimes l} )=S(\cn^{\otimes l}(\pi_{\delta,l,\cn}^{(t)} ) )- S(\cn_c^{\otimes l}(\pi_{\delta,l,\cn}^{(t)} ) ), \]
the inequalities (\ref{eq:compound-bsst-ie-lemma-3}),(\ref{eq:compound-bsst-ie-lemma-4}), (\ref{eq:compound-bsst-ie-lemma-7}), (\ref{eq:compound-bsst-ie-lemma-7a}), (\ref{eq:compound-bsst-ie-lemma-8}), (\ref{eq:compound-bsst-ie-lemma-9}) and triangle inequality show that \emph{uniformly} in $\cn\in\fri$ we have
\begin{eqnarray}\label{eq:compound-bsst-ie-lemma-10}
  \left |\frac{1}{l}I_c(\pi_{\delta,l,\cn}^{(t)},\cn^{\otimes l} )- I_c(\rho_{\cn}^{(t)},\cn )   \right|&\le & \theta_l\left(\delta,\frac{t}{\dim\hr},\frac{t}{\dim \kr}\right)+ \theta_l\left(\delta,\frac{t}{\dim\hr},\frac{t}{\dim \hr_e}\right)\nonumber\\
&&- t \log\frac{t}{\dim\kr\cdot \dim\hr_e} -l\cdot t\log \frac{l\cdot t}{\dim\kr\cdot \hr_e}.
\end{eqnarray}
Now, by (\ref{eq:def-perturbed-state}) we have
\[||\rho_{\cn}^{(t)}-\rho_{\cn}   ||_1\le t  \]
which implies
\[||\cn(\rho_{\cn}^{(t)} )-\cn (\rho_{\cn} )||_1\le t,\qquad ||\cn_c(\rho_{\cn}^{(t)} )-\cn_c (\rho_{\cn} )||_1\le t,   \]
since the trace distance of two states can only decrease after applying a trace preserving  completely positive map to both states. 
Thus Fannes inequality leads us to the conclusion that 
\[  \left|I_c(\rho_{\cn}^{(t)},\cn )-I_c(\rho_{\cn},\cn )   \right|\le -t\log \frac{t}{\dim\kr\cdot \dim\hr_e}.  \]
This and (\ref{eq:compound-bsst-ie-lemma-10}) shows that uniformly in $\cn\in \fri$
\begin{eqnarray}\label{eq:compound-bsst-ie-lemma-11}
  \left |\frac{1}{l}I_c(\pi_{\delta,l,\cn}^{(t)},\cn^{\otimes l} )- I_c(\rho_{\cn},\cn )   \right|&\le & \theta_l\left(\delta,\frac{t}{\dim\hr},\frac{t}{\dim \kr}\right)+ \theta_l\left(\delta,\frac{t}{\dim\hr},\frac{t}{\dim \hr_e}\right)\nonumber\\
&&- 2t \log\frac{t}{\dim\kr\cdot \dim\hr_e} -l\cdot t\log \frac{l\cdot t}{\dim\kr\cdot \hr_e}\nonumber\\
&=:& \frac{\Delta_l(\delta,t)}{2}.
\end{eqnarray}
Finally, it is clear from the uniform estimate in (\ref{eq:compound-bsst-ie-lemma-11}) that
\begin{eqnarray}\label{eq:compound-bsst-ie-lemma-12}
  \left|\frac{1}{l}\inf_{\cn\in\fri}I_c(\pi_{\delta,l,\cn}^{(t)},\cn^{\otimes l} )-\inf_{\cn\in\fri}\max_{\rho\in\cs(\hr)}I_c(\rho,\cn)      \right|&=&\left |\frac{1}{l}\inf_{\cn\in\fri}I_c(\pi_{\delta,l,\cn}^{(t)},\cn^{\otimes l} )- \inf_{\cn\in\fri}I_c(\rho_{\cn},\cn )   \right|\nonumber\\
&\le & \Delta_l(\delta,t),
\end{eqnarray}
which concludes the proof.
 \begin{flushright}$\Box$\end{flushright}
Lemma \ref{compound-bsst-ie-lemma} and Theorem \ref{direct-finite-informed-encoder} easily imply the following result.
\begin{lemma}\label{general-ie-direct-part}
Let $\fri\subset\mathcal{C}(\hr,\kr)$ be an arbitrary set of quantum channels. Then
\[Q_{IE}(\fri)\ge \inf_{\cn\in\fri}\max_{\rho\in\mathcal{S}(\hr)}I_c(\rho,\cn).   \]
\end{lemma}
\emph{Proof.}
Take any set $\{\pi_{\gr_\cn}\}_{\cn\in\fri}$ of maximally mixed states on subspaces $\gr_\cn\subset\hr$. In a first step we will show that
\begin{equation}\label{eq:direct-general-informed-encoder-0}
 Q_{IE}(\fri)\ge \inf_{\cn\in \fri}I_c(\pi_{\gr_\cn},\cn) 
\end{equation}
holds. Notice that we can assume w.l.o.g. that $\inf_{\cn\in \fri}I_c(\pi_{\gr_\cn},\cn)>0 $.\\
 Denote, for every $\tau>0$, by $\fri_\tau$ a $\tau$-net for $\fri$ as given in (\ref{adapted-net}) of cardinality $N_\tau:=|\fri_\tau|\leq(\frac{6}{\tau})^{2(d\cdot d')^2}$, where $d,d'$ are the dimensions of $\hr,\kr$. Starting from this set $\fri_{\tau}$ it is easy to construct a finite set $\fri_{\tau}^{\circ}$ with the following properties:
 \begin{enumerate}
 \item $\fri_{\tau}^{\circ}\subset \fri $,
 \item $|\fri_{\tau}^{\circ}|\leq(\frac{6}{\tau})^{2(d\cdot d')^2} $, and
 \item to each $\cn\in \fri$ there is at least one $\cn'\in\fri_{\tau}^{\circ}$ with $||\cn-\cn'||_{\lozenge}\le 2\tau$.
 \end{enumerate}
Let $(\tau_l)_{l\in\mathbb N}$ be defined by $\tau_l:=\frac{1}{l^2}$ and consider the sets $\fri_{\tau_l}^{\circ}$, $l\in\nn$.\\
Take any $\eta\in (0, \inf_{\cn\in \fri}I_c(\pi_{\gr_\cn},\cn))$ and set 
\[R(\eta):=\inf_{\cn\in\fri}I_c(\pi_{\gr_\cn},\cn)-\eta,\]
and
\[R_l(\eta):=\min_{\cn\in\fri_{\tau_l}^{\circ}}I_c(\pi_{\gr_\cn},\cn)-\eta.  \] 
Then for every $l\in\nn$, 
\begin{equation}\label{eq:direct-general-informed-encoder-00}
 R_l(\eta)\ge R(\eta)
\end{equation} 
since $\fri_{\tau_l}^{\circ}\subset \fri$.\\
Fix some $\delta'\in(0,1/2)$ such that $\gamma(\delta')+\vphi(\delta')<\eta/4$. For every $l\in\mathbb N$, choose a subspace $\E_l\subset\hr^{\otimes l}$ of dimension 
\[k_l(\eta):=\dim\E_l=\lfloor2^{lR_l(\eta)}\rfloor .\] 
The proof of Theorem \ref{direct-finite-informed-encoder} then shows the existence of a recovery operation $\crr^l$ and for each $\cn'\in\fri_{\tau_l}^{\circ}$ a unitary encoder $\mathcal{W}_{\cn'}^{l}$ such that for each $l\in\nn$
$$F_e(\pi_{\E_l},\crr^l\circ\cn'^{\otimes l}\circ\mathcal{W}_{\cn'}^{l})\geq1-3\cdot N_{\tau_l}\cdot\eps_l\qquad \forall\ \cn'\in\fri_{\tau_l}^{\circ},$$
where $\eps_l:=2^{-l(c\delta'^2-h(l))}+2^{-l(c'\delta'^2-h'(l))} +2 N_{\tau_l}\sqrt{2^{l(-\frac{3\eta}{4}+h'(l))}})$. From Lemma \ref{lemma-approx-properties} along  with the properties of $\fri_{\tau_l}^{\circ}$ and our specific choice of $(\tau_l)_{l\in\mathbb N}$ it follows that there exist unitary encodings $\mathcal W^l_\cn$ (for every $l\in\mathbb N$ and each $\cn\in\fri$), such that
$$F_e(\pi_{\E_l},\crr^l\circ\cn^{\otimes l}\circ\mathcal W^l_\cn)\geq1-3\cdot N_{\tau_l}\cdot\eps_l-\frac{2}{l}\ \forall l\in\mathbb N,\ \cn\in\fri.$$
Clearly, $\lim_{l\rightarrow\infty}F_e(\pi_{\E_l},\crr^l\circ\cn^{\otimes l}\circ\mathcal W^l_\cn)=1$ and (\ref{eq:direct-general-informed-encoder-00}) implies for each $\eta\in (0, \inf_{\cn\in \fri}I_c(\pi_{\gr_\cn},\cn))$ that 
\[ \liminf_{l\to\infty}\frac{1}{l}\log k_l(\eta)=\liminf_{l\to\infty}\frac{1}{l}\log\dim\E_l  \ge R(\eta).\] 
Consequently $\inf_{\cn\in\fri}I_c(\pi_{\gr_\cn},\cn)$ is achievable.\\
We proceed by repeated application of the inequality 
\begin{equation}\label{eq:direct-general-informed-encoder-2}
Q_{IE}(\fri)\ge\frac{1}{l}Q_{IE}(\fri^{\otimes l})\ (\forall l\in\mathbb N).
\end{equation}
From (\ref{eq:direct-general-informed-encoder-0}) and (\ref{eq:direct-general-informed-encoder-2}) we get that for each $l\in\mathbb N$ and every set $\{\pi_{\cn}^l\}_{\cn\in\fri}$ of maximally mixed states on subspaces of $\hr^{\otimes l}$, 
$$Q_{IE}(\fri)\geq\frac{1}{l}\inf_{\cn\in\fri}I_c(\pi^l_\cn,\cn^{\otimes l}).$$
We now make a specific choice of the states $\pi^l_\cn$, namely, for every $\cn\in\fri$ and $l\in\mathbb N$, set $\pi^l_\cn:=\pi_{\delta_l,l,\cn}^{(t_l)}$ with $\pi_{\delta_l,l,\cn}^{(t_l)}$ taken from the second part of Lemma \ref{compound-bsst-ie-lemma}. By an application of the second part of Lemma \ref{compound-bsst-ie-lemma} it follows
\begin{eqnarray*}
Q_{IE}(\fri)&\geq&\lim_{l\rightarrow\infty}\frac{1}{l}\inf_{\cn\in\fri}I_c(\pi^l_\cn,\cn^{\otimes l})\\
&\geq&\lim_{l\rightarrow\infty}(\inf_{\cn\in\fri}I_c(\rho_\cn,\cn)-\nu_l)\\
&=&\inf_{\cn\in\fri}I_c(\rho_\cn,\cn)\\
&=&\inf_{\cn\in\fri}\max_{\rho\in\cs(\hr)}I_c(\rho,\cn).
\end{eqnarray*}
\begin{flushright}$\Box$\end{flushright}
Employing inequality (\ref{eq:direct-general-informed-encoder-2}) one more time we obtain from Lemma \ref{general-ie-direct-part} applied to $\fri^{\otimes l}$
\begin{eqnarray*}
  Q_{IE}(\fri)&\ge& \frac{1}{l}Q_{IE}(\fri^{\otimes l})\\
              &\ge&  \frac{1}{l}\inf_{\cn\in\fri}\max_{\rho\in \mathcal S(\hr^{\otimes l})}I_c(\rho, \cn^{\otimes l}).
\end{eqnarray*}
Consequently we obtain the desired achievability result.
\begin{theorem}[Direct Part: Informed Encoder]\label{final-direct-part-ie}
For any $\fri\in\mathcal{C}(\hr,\kr)$ we have
\begin{equation}\label{eq:final-direct-part-ie}
 Q_{IE}(\fri)\ge \lim_{l\to\infty}\frac{1}{l}\inf_{\cn\in\fri}\max_{\rho\in \mathcal S(\hr^{\otimes l})}I_c(\rho, \cn^{\otimes l}).
\end{equation}
\end{theorem}
\begin{remark}
Note that the limit in (\ref{eq:final-direct-part-ie}) exists. Indeed, set
\[ C_l(\cn):= \max_{\rho\in\mathcal{S}(\hr^{\otimes l})}I_c(\rho,\cn^{\otimes l}).  \]
Then it is clear that
 \[ C_{l+k}(\cn)\ge C_l(\cn)+C_k(\cn)\]
and consequently
\begin{eqnarray*}
  \inf_{\cn\in\fri}C_{l+k}(\cn)&\ge& \inf_{\cn\in\fri}(C_l(\cn)+C_k(\cn))\\
                              &\ge& \inf_{\cn\in\fri}C_l(\cn)+\inf_{\cn\in\fri}C_k(\cn),
\end{eqnarray*}
which implies the existence of the limit in (\ref{eq:final-direct-part-ie}).
\end{remark}
\section{\label{sec:Converse Parts of the Coding Theorems for General Quantum Compound Channels}Converse Parts of the Coding Theorems for General Quantum Compound Channels}
In this section we prove the converse parts of the coding theorems for general quantum compound channels in the three different settings concerned with entanglement transmission that are treated in this paper. The proofs deviate from the usual approach due to our more general definitions of codes.
\subsection{\label{subsec:Converse:The Case of Uninformed Users}Converse for Informed Decoder and Uninformed Users}
We first prove the converse part in the case of a finite compound channel, then use a recent result \cite{leung-smith} that gives a more convenient estimate for the difference in coherent information of two nearby channels in order to pass on to the general case.\\
For the converse part in the case of a finite compound channel we need the following lemma that is due to Devetak  \cite{devetak}:
\begin{lemma}[Cf. \cite{devetak}]\label{lemma:converse-uninformed-devetak}
For two states $\sigma,\rho\in\mathcal S(\hr_1\otimes\hr_2)$ where $\dim\hr_1\otimes\hr_2=b$ with fidelity $f=F(\sigma,\rho)$,
$$|\Delta S(\rho)-\Delta S(\sigma)|\leq\frac{2}{e}+4\log(b)\sqrt{1-f},$$
where
$$\Delta S(\ \cdot\ ):=S(\tr_{\hr_1}[\ \cdot\ ])-S(\ \cdot\ ).$$
\end{lemma}
We shall now embark on the proof of the following theorem.
\begin{theorem}[Converse Part: Informed Decoder, Uninformed Users, $|\fri|<\infty$]\label{theorem:converse-uninformed-finite}
Let $\fri=\{\cn_1,\ldots,\cn_N\}\subset\mathcal C(\hr,\kr)$ be a finite compound channel. The capacities $Q_{ID}(\fri)$ and $Q(\fri)$ of $\fri$ with informed decoder and uninformed users are bounded from above by
$$Q(\fri)\leq Q_{ID}(\fri)\leq\lim_{l\rightarrow\infty}\max_{\rho\in\mathcal S(\hr^{\otimes l})}\min_{\cn_i\in\fri}\frac{1}{l}I_c(\rho,\cn_i^{\otimes l}).$$
\end{theorem}
\emph{Proof.} The inequality $Q(\fri)\leq Q(\fri)_{ID}$ is obvious from the definition of codes. We give a proof for the second inequality. Let for arbitrary $l\in\mathbb N$ an $(l,k_l)$ code for a compound channel $\fri=\{\cn_1,\ldots,\cn_N\}$ with informed decoder and the property $\min_{1\leq i\leq N}F_e(\pi_{\fr_l},\crr^l_i\circ\cn_i^{\otimes l}\circ\mathcal P^l)\geq1-\epsilon_l$ be given, where $\epsilon_l\in[0,1]$. Let $|\psi_l\rangle\langle\psi_l|\in\mathcal S(\E_l\otimes\fr_l)$ be a purification of $\pi_{\fr_l}$ where $\E_l$ is just a copy of $\fr_l$. We use the abbreviation $\D^l:=\frac{1}{N}\sum_{i=1}^N\crr^l_i\circ\cn_i^{\otimes l}$. Obviously, the above code then satisfies
\begin{eqnarray}
\langle\psi_l,id_{\E_l}\otimes\D^l(id_{\E_l}\otimes\mathcal P^l(|\psi_l\rangle\langle\psi_l|))\psi_l\rangle
&=&\frac{1}{N}\sum_{i=1}^NF_e(\pi_{\fr_l},\crr_i^l\circ\cn_i^{\otimes l}\circ\mathcal P^l)\nonumber\\
&\geq&1-\epsilon_l.\label{eq:converse-uninformed-1}
\end{eqnarray}
Let $\sigma_{\mathcal P^l}:=id_{\E_l}\otimes\mathcal P^l(|\psi^l\rangle\langle\psi^l|)$ and consider any convex decomposition $\sigma_{\mathcal P^l}=\sum_{i=1}^{(\dim\fr_l)^2}\lambda_i|e_i\rangle\langle e_i|$ of $\sigma_{\mathcal P^l}$ into pure states $|e_i\rangle\langle e_i|\in\mathcal S(\fr_l\otimes\hr^{\otimes l})$. By (\ref{eq:converse-uninformed-1}) there is at least one $i\in\{1,\ldots,(\dim\fr_l)^2\}$ such that
\begin{equation}\langle\psi_l,id_{\E_l}\otimes\D^l(|e_i\rangle\langle e_i|)\psi_l\rangle\geq1-\epsilon_l\label{eq:converse-uninformed-2}\end{equation}
holds. Without loss of generality, $i=1$. Turning back to the individual channels, we get
\begin{equation}\langle\psi_l,id_{\E_l}\otimes\crr_i^l\circ\cn_i^{\otimes l}(|e_1\rangle\langle e_1|)\psi_l\rangle\geq1-N\epsilon_l\ \ \ \forall i\in\{1,\ldots,N\}\label{eq:converse-uninformed-3}.\end{equation}
We define the state $\rho^l:=\tr_{\E_l}(|e_1\rangle\langle e_1|)\in\mathcal S(\hr^{\otimes l})$ and note that $|e_1\rangle\langle e_1|$ is a purification of $\rho^l$.
Application of recovery operation and individual channels to $\rho^l$ now defines the states $\sigma^l_k:=id_{\E_l}\otimes\crr_k^l\circ\cn_k^{\otimes l}(|e_1\rangle\langle e_1|)$ ($k\in\{1,\ldots,N\}$) which have independently of $k$ the property
$$F(\psi^l,\sigma^l_k)=\langle\psi_l,id_{\E_l}\otimes\crr_k^l\circ\cn_k^{\otimes l}(|e_i\rangle\langle e_i|)\psi_l\rangle\geq1-N\epsilon_l$$
and thus put us into position for an application of Lemma \ref{lemma:converse-uninformed-devetak}, which together with the data processing inequality for coherent information \cite{schumacher-nielsen} establishes the following chain of inequalities for every $k\in\{1,\ldots,N\}$:
\begin{eqnarray}
\log\dim\fr_l&=&S(\pi_{\fr_l})\nonumber\\
&=&\Delta S(|\psi^l\rangle\langle\psi^l|)\nonumber\\
&\leq &\Delta S(\sigma_k^l)+\frac{2}{e}+4\log((\dim\fr_l)^2)\sqrt{N\epsilon_l}\nonumber\\
&=&S(\tr_{\E_l}(id_{\E_l}\otimes\crr_k^l\circ\cn^{\otimes l}(|e_1\rangle\langle e_1|))-S(id_{\E_l}\otimes\crr_k^l\circ\cn_k^{\otimes l}(|e_1\rangle\langle e_1|))\nonumber\\
&&+\frac{2}{e}+4\log((\dim\fr_l)^2)\sqrt{N\epsilon_l}\nonumber\\
&=&I_c(\rho^l,\crr_k^l\circ\cn_k^{\otimes l})+\frac{2}{e}+4\log((\dim\fr_l)^2)\sqrt{N\epsilon_l}\nonumber\\
&\leq&I_c(\rho^l,\cn_k^{\otimes l})+\frac{2}{e}+4\log((\dim\fr_l)^2)\sqrt{N\epsilon_l}.\label{eq:converse-uninformed-5}
\end{eqnarray}
Thus,
\begin{eqnarray}
\log\dim\fr_l&\leq&\min_{k\in\{1,\ldots,N\}}I_c(\rho^l,\cn_k^{\otimes l})+\frac{2}{e}+4\log((\dim\fr_l)^2)\sqrt{N\epsilon_l}\nonumber\\
&\leq&\max_{\rho\in\mathcal S(\hr^{\otimes l})}\min_{k\in\{1,\ldots,N\}}I_c(\rho,\cn_k^{\otimes l})+\frac{2}{e}+8\log(\dim\fr_l)\sqrt{N\epsilon_l}\label{eq:converse-uninformed-4}.
\end{eqnarray}
Let a sequence of $(l,k_l)$ codes for $\fri$ with informed decoder be given such that $\liminf_{l\rightarrow\infty}\frac{1}{l}\log\dim\fr_l=R\in\mathbb R$ and $\lim_{l\rightarrow\infty}\epsilon_l=0$. Then by (\ref{eq:converse-uninformed-4}) we get
\begin{eqnarray*}
R&=&\liminf_{l\rightarrow\infty}\frac{1}{l}\log\dim\fr_l\\
&\leq&\liminf_{l\rightarrow\infty}\frac{1}{l}\max_{\rho\in\mathcal S(\hr^{\otimes l})}\min_{k\in\{1,\ldots,N\}}I_c(\rho,\cn_k^{\otimes l})\\
&&+\liminf_{l\rightarrow\infty}\frac{1}{l}\frac{2}{e}+\liminf_{l\rightarrow\infty}8\log(\dim\fr_l)\sqrt{N\epsilon_l}\\
&=&\lim_{l\rightarrow\infty}\frac{1}{l}\max_{\rho\in\mathcal S(\hr^{\otimes l})}\min_{k\in\{1,\ldots,N\}}I_c(\rho,\cn_k^{\otimes l}),
\end{eqnarray*}
\begin{flushright}$\Box$\end{flushright}
Let us now focus on the general case. We shall prove the following theorem:
\begin{theorem}[Converse Part: Informed Decoder, Uninformed Users]\label{theorem:converse-uninformed-general}
Let $\fri\subset\mathcal C(\hr,\kr)$ be a compound channel. The capacities $Q_{ID}(\fri)$ and $Q(\fri)$ for $\fri$ with informed decoder and with uninformed users are bounded from above by
$$Q(\fri)\leq Q_{ID}(\fri)\leq\lim_{l\rightarrow\infty}\max_{\rho\in\mathcal S(\hr^{\otimes l})}\inf_{\cn\in\fri}\frac{1}{l}I_c(\rho,\cn^{\otimes l}).$$
\end{theorem}
For the proof of this theorem, we will make use of the following Lemma:
\begin{lemma}[Cf. \cite{leung-smith}]\label{lemma:entropy-estimate-of-leung&smith}
Let $\cn,\cn_i\in\mathcal C(\hr,\kr)$ and $d_\kr=\dim\kr$. Let $\hr_r$ be an additional Hilbert space , $l\in\mathbb N$ and $\phi\in\mathcal S(\hr_r\otimes\hr^{\otimes l})$. If $||\cn-\cn_i||_{\lozenge}\leq\epsilon$, then
$$|S(id_{\hr_r}\otimes\cn^{\otimes l}(\phi))-S(id_{\hr_r}\otimes\cn_i^{\otimes l}(\phi))|\leq l(4\epsilon\log(d_\kr)+2h(\epsilon)).$$
Here, $h(\cdot)$ denotes the binary entropy.
\end{lemma}
This result immediately implies the following Lemma:
\begin{lemma}\label{lemma:estimate-for-coherent-information}
Let $\hr,\kr$ be finite dimensional Hilbert spaces. There is a function $\nu:[0,1]\rightarrow\mathbb R_+$ with $\lim_{x\rightarrow0}\nu(x)=0$ such that for every $\fri,\fri'\subseteq \mathcal{C}(\hr,\kr) $ with $D_\lozenge(\fri,\fri')\leq\tau\leq1/2$ and every $l\in\mathbb N$ we have the estimates
\begin{enumerate}
\item $$|\frac{1}{l}I_c(\rho,\fri^{\otimes l})-\frac{1}{l}I_c(\rho,\fri'^{\otimes l})|\leq\nu(2\tau)\ \ \ \forall \rho\in\mathcal S(\hr^{\otimes l})$$
\item $$|\frac{1}{l}\inf_{\cn\in\fri}\max_{\rho\in\mathcal S(\hr^{\otimes l})}I_c(\rho,\cn^{\otimes l})-\frac{1}{l}\inf_{\cn'\in\fri'}\max_{\rho\in\mathcal S(\hr^{\otimes l})}I_c(\rho,\cn'^{\otimes l})|\leq\nu(2\tau)$$
\end{enumerate}
The function $\nu$ is given by $\nu(x)=x+8x\log(d_\kr)+4h(x)$. Again, $h(\cdot)$ denotes the binary entropy.
\end{lemma}
\emph{Proof of Theorem \ref{theorem:converse-uninformed-general}.} Again, the first inequality is easily seen to be true from the very definition of codes in the two cases, so we concentrate on the second. Let $\fri\subset\mathcal C(\hr,\kr)$ be a compound channel and let for every $\l\in\mathbb N$ an $(l,k_l)$ code for $\fri$ with informed decoder be given such that
$\liminf_{l\to\infty}\frac{1}{l}\log k_l=R$, and $\lim_{l\to\infty}\inf_{\cn\in\fri}F_e(\pi_{\fr_l},\crr_{\cn}^l\circ \cn^{\otimes l}\circ\mathcal{P}^l)=1 $ hold.\\
Take any $0<\tau\le1/2$. Then it is easily seen that starting with a $\frac{\tau}{2}$-net in $\mathcal{C}(\hr,\kr)$ we can find a set $\fri'_{\tau}=\{\cn_1,\ldots,\cn_{N_{\tau}}  \}\subset\fri$ with $|N_{\tau}|\le (\frac{6}{\tau})^{2(\dim\hr \cdot \dim \kr)^2}$ such that for each $\cn\in\fri$ there is $\cn_i\in\fri'_{\tau}$ with
\[||\cn-\cn_i||_{\lozenge}\le \tau.  \]
Clearly, the above sequence of codes satisfies for each $i\in\{  1,\ldots, N_{\tau}\}$
\begin{enumerate}
\item $\liminf_{l\to\infty}\frac{1}{l}\log k_l=R$, and \item
$\lim_{l\to\infty}\min_{\cn_i\in\fri_\tau}F_e(\pi_{\fr_l},\crr^l\circ \cn_i^{\otimes l}\circ\mathcal{P}^l)=1 $.
\end{enumerate}
From Theorem \ref{theorem:converse-uninformed-finite} it is immediately clear then, that
$$R\leq \lim_{l\rightarrow\infty}\max_{\rho\in\mathcal S(\hr^{\otimes l})}\min_{\cn_i\in\fri'_{\tau}}\frac{1}{l}I_c(\rho,\cn_i^{\otimes l})$$
and from the first estimate in Lemma \ref{lemma:estimate-for-coherent-information} we get by noting that $D_\lozenge(\fri,\fri'_\tau)\leq\tau$ holds
$$R\leq \lim_{l\rightarrow\infty}\max_{\rho\in\mathcal S(\hr^{\otimes l})}\inf_{\cn\in\fri}\frac{1}{l}I_c(\rho,\cn^{\otimes l})+\nu(2\tau).$$
Taking the limit $\tau\rightarrow0$ proves the theorem.
\subsection{\label{subsec:Converse:The Informed Encoder}The Informed Encoder}
The case of an informed encoder can be treated in the same manner as the other two cases. We will just state the theorem and very briefly indicate the central ideas of the proof.
\begin{theorem}[Converse Part: Informed Encoder]\label{theorem:converse-informed-encoder-general}
Let $\fri\subset\mathcal C(\hr,\kr)$ be a compound channel. The capacity $Q_{IE}(\fri)$ for $\fri$ with informed encoder is bounded from above by
$$Q_{IE}(\fri)\leq\lim_{l\rightarrow\infty}\frac{1}{l}\inf_{\cn\in\fri}\max_{\rho\in\mathcal S(\hr^{\otimes l})}I_c(\rho,\cn^{\otimes l}).$$
\end{theorem}
\emph{Proof.} The proof of this theorem is a trivial modification of the one for Theorem \ref{theorem:converse-uninformed-general}. Again, the first part of the proof is the converse in the finite case, while the second part uses the second estimate in Lemma \ref{lemma:estimate-for-coherent-information}.\\
For the proof in the finite case note the following: due to the data processing inequality, the structure of the proof is entirely independent from the decoder. A change from an informed decoder to an uninformed decoder does not change our estimate. The only important change is that there will be a whole set $\{e^1_{i_1},\ldots,e^N_{i_N}\}$ of vector states satisfying equation (\ref{eq:converse-uninformed-2}), one for each channel in $\fri$. This causes the state $\rho^l$ in equation (\ref{eq:converse-uninformed-5}) to depend on the channel.
\begin{flushright}$\Box$\end{flushright}
\section{\label{sec:Continuity of Compound Capacity}Continuity of Compound Capacity}
This section is devoted to a question that has been answered only recently in \cite{leung-smith} for single-channel capacities, namely that of continuity of capacities of quantum channels.\\
The question is relevant not only from a mathematical point of view, but might also have a strong impact on applications. It seems a hard task in general to compute the regularized capacity formulas obtained so far for quantum channels. There are, however, cases where the regularized capacity formula can be reduced to a one-shot quantity (see for example \cite{cubitt-ruskai-smith} and references therein) that can be calculated using standard optimization techniques.\\
Knowing that capacity is a continuous quantity one could raise the question how close an arbitrary (compound) channel is to a (compound) channel with one-shot capacity and thereby get an estimate on arbitrary capacities.\\
We will now state the main result of this section.
\begin{theorem}[Continuity of Compound Capacity]\label{theorem:continuity-of-compound-capacity}
The compound capacities $Q(\ \cdot\ ),\ Q_{ID}(\ \cdot\ )$ and $Q_{IE}(\ \cdot\ )$ are continuous. To be more precise, let $\fri,\fri'\subset\mathcal C(\hr,\kr)$ be two compound channels with $D_\lozenge(\fri,\fri')\leq\epsilon\leq1/2$. Then $$|Q(\fri)-Q(\fri')|=|Q_{ID}(\fri)-Q_{ID}(\fri')|\leq\nu(2\epsilon),$$
$$|Q_{IE}(\fri)-Q_{IE}(\fri')|\leq\nu(2\epsilon),$$
where the function $\nu$ is taken from Lemma \ref{lemma:estimate-for-coherent-information}.
\end{theorem}
\begin{remark}
Let $\fri\subset\mathcal C(\hr,\kr)$. Then $D(\fri,\bar\fri)=0$, implying that the three different capacities of $\fri$ coincide with those for $\bar\fri$. We may thus define the equivalence relation $\fri\sim\fri'\Leftrightarrow\bar\fri=\bar{\fri'}$ and even use $D_\lozenge$ as a metric on the set of equivalence classes without losing any information about our channels.
\end{remark}
\emph{Proof.} Let $D_\lozenge(\fri,\fri')\leq\epsilon$. By the first estimate in Lemma \ref{lemma:estimate-for-coherent-information} and the capacity formula $Q_{ID}(\fri)=Q(\fri)=\lim_{l\rightarrow\infty}\frac{1}{l}\max_{\rho\in\mathcal S(\hr^{\otimes l})}I_c(\rho,\fri^{\otimes l})$ we get
\begin{eqnarray*}
|Q(\fri)-Q(\fri')|&=&|Q_{ID}(\fri)-Q_{ID}(\fri')|\\
&=&|\lim_{l\rightarrow\infty}\frac{1}{l}[\max_{\rho\in\mathcal S(\hr^{\otimes l})}I_c(\rho,\fri^{\otimes l})-\max_{\rho\in\mathcal S(\hr^{\otimes l})}I_c(\rho,\fri'^{\otimes l})]|\\
&=&\lim_{l\rightarrow\infty}|\frac{1}{l}\max_{\rho\in\mathcal S(\hr^{\otimes l})}I_c(\rho,\fri^{\otimes l})-\frac{1}{l}\max_{\rho\in\mathcal S(\hr^{\otimes l})}I_c(\rho,\fri'^{\otimes l})|\\
&\leq&\lim_{l\rightarrow\infty}\nu(2\epsilon)\\
&=&\nu(2\epsilon).
\end{eqnarray*}
For the proof in the case of an informed encoder let us first note
that
$Q_{IE}(\fri)=\lim_{l\rightarrow\infty}\inf_{\cn\in\fri}\max_{\rho\in\mathcal
S(\hr^{\otimes l})}I_c(\rho,\cn^{\otimes l})$ holds. The second
estimate in Lemma \ref{lemma:estimate-for-coherent-information}
justifies the following inequality:
\begin{eqnarray*}
|Q_{IE}(\fri)-Q_{IE}(\fri')|&=&|\lim_{l\rightarrow\infty}\frac{1}{l}[\inf_{\cn\in\fri}\max_{\rho\in\mathcal S(\hr^{\otimes l})}I_c(\rho,\cn^{\otimes l})-\inf_{\cn'\in\fri}\max_{\rho\in\mathcal S(\hr^{\otimes l})}I_c(\rho,\cn'^{\otimes l})]|\\
&=&\lim_{l\rightarrow\infty}|\frac{1}{l}\inf_{\cn\in\fri}\max_{\rho\in\mathcal S(\hr^{\otimes l})}I_c(\rho,\cn^{\otimes l})-\frac{1}{l}\inf_{\cn'\in\fri}\max_{\rho\in\mathcal S(\hr^{\otimes l})}I_c(\rho,\cn'^{\otimes l})|\\
&\leq&\lim_{l\rightarrow\infty}\nu(2\epsilon)\\
&=&\nu(2\epsilon).
\end{eqnarray*}
\begin{flushright}$\Box$\end{flushright}
\section{\label{sec:ent-generating} Entanglement-Generating Capacity of Compound Channels}
In this last section we will use the results obtained so far to achieve our main goal. Namely, we will determine the entanglement-generating capacity of quantum compound channels. We give the definitions of codes and capacity only for the most interesting case of uninformed users because there is no doubt that the reader will easily guess the definitions in the remaining cases. Nevertheless, we will state the coding result in all three cases.\\
An entanglement-generating $(l,k_l)$-code for the compound channel $\fri\subset\mathcal{C}(\hr,\kr)$ with uninformed users consists of a pair $(\crr^l, \varphi_l)$ where $\crr^l\in\mathcal{C}(\kr^{\otimes l},\fr_l)$ with $k_l=\dim \fr_l$ and $\varphi_l$ is a pure state on $\fr_l\otimes \hr^{\otimes l}$.\\
$R\in\rr_+$ is called an achievable rate for $\fri$ with uninformed users if there is a sequence of $(l,k_l)$ entanglement-generating codes with
\begin{enumerate}
\item $\liminf_{l\to\infty}\frac{1}{l}\log k_l\ge R$, and
\item $\lim_{l\to\infty}\inf_{\cn\in\fri}F(|\psi_{l}\rangle\langle \psi_{l}|, (id_{\fr_l}\otimes \crr^{l}\circ \cn^{\otimes l})(|\varphi_l\rangle\langle \varphi_l| ))=1$ where $\psi_{l}$ denotes the standard maximally entangled state on $\fr_l\otimes\fr_l$ and $F(\cdot ,\cdot)$ is the fidelity.
\end{enumerate}
The entanglement-generating capacity of $\fri$ with uninformed users is then defined as the least upper bound of all achievable rates and is denoted by $E(\fri)$. The entanglement-generating capacities $E_{ID}(\fri)$ and $E_{IE}(\fri)$ of $\fri$ with informed decoder or informed encoder are obtained if we allow the decoder or preparator to choose $\crr^l$ or $\varphi_l$ in dependence of $\cn\in\fri$.\\
Recall from the proof of Theorem \ref{direct-coding-general-uu} that to each subspace $\gr\subset \hr$ and $\epsilon>0$ we always can find a subspace $\fr_l\subset\gr^{\otimes l}\subset\hr^{\otimes l}$, a recovery operation $\crr^l\in \mathcal{C}(\kr^{\otimes l},\fr_l)$, and a unitary operation $\U^l\in\mathcal{C}(\hr^{\otimes l},\hr^{\otimes l})$ with
\begin{equation}\label{eq:ent-gen-0}
 k_l=\dim \fr_l\ge \lfloor 2^{l ( \inf_{\cn\in \fri} I_c(\pi_{\gr},\cn )-\frac{\epsilon}{2}-o(l^0)) }  \rfloor,
\end{equation}
and
\begin{equation}\label{eq:ent-gen-1}
  \inf_{\cn\in\fri}F_e(\pi_{\fr_l}, \crr^l\circ\cn^{\otimes l}\circ \U^l )= 1-o(l^0).
\end{equation}
Notice that the maximally entangled state $\psi_l$ in $\fr_l\otimes \fr_l$ purifies the maximally mixed state $\pi_{\fr_l}$on $\fr_l$ and defining $|\varphi_l\rangle\langle \varphi_l|:= \U^l(|\psi_{l}\rangle\langle \psi_{l}|  )$, the relation (\ref{eq:ent-gen-1}) can be rewritten as
\begin{equation}\label{eq:ent-gen-2}
 \inf_{\cn\in\fri}F(|\psi_{l}\rangle\langle \psi_{l}|, id_{\fr_l}\otimes \crr^{l}\circ \cn^{\otimes l}(|\varphi_l\rangle\langle \varphi_l| ))=1-o(l^0).
\end{equation}
This together with (\ref{eq:ent-gen-0}) shows that
\begin{equation}\label{eq:ent-gen-3}
  E(\fri)\ge \inf_{\cn\in \fri} I_c(\pi_{\gr},\cn ).
\end{equation}
Thus, using the compound BSST Lemma \ref{compound-bsst-lemma} and arguing as in the proof of Theorem \ref{final-direct-part-uu}, we can conclude that
\begin{equation}\label{eq:ent-gen-4}
  E(\fri)\ge Q(\fri)= \lim_{l\to\infty}\frac{1}{l} \max_{\rho\in\mathcal{S}(\hr^{\otimes l})}\inf_{\cn\in\fri}I_c(\rho,\cn^{\otimes l}).
\end{equation}
Since $E(\fri)\le E_{ID}(\fri)$ holds it suffices to show
\begin{equation}\label{eq:ent-gen-5}
E_{ID}(\fri)\le Q(\fri)= \lim_{l\to\infty}\frac{1}{l} \max_{\rho\in\mathcal{S}(\hr^{\otimes l})}\inf_{\cn\in\fri}I_c(\rho,\cn^{\otimes l})
\end{equation}
in order to establish the coding theorem for $E_{ID}(\fri)$ and $E(\fri)$ simultaneously.\\
The proof of (\ref{eq:ent-gen-5}) relies on Lemma \ref{lemma:converse-uninformed-devetak} and the data processing inequality. Indeed, let $R\in\rr_+$ be an achievable entanglement generation rate for $\fri$ with informed decoder and let $((\crr_{\cn}^l)_{\cn\in\fri}, \varphi_l)_{l\in\nn}$ be a corresponding sequence of $(l,k_l)$-codes, i.e we have
\begin{enumerate}
\item $\liminf_{l\to\infty}\frac{1}{l}\log k_l\ge R$, and
\item $\inf_{\cn\in\fri}F(|\psi_{l}\rangle\langle \psi_{l}|, (id_{\fr_l}\otimes \crr_{\cn}^{l}\circ \cn^{\otimes l})(|\varphi_l\rangle\langle \varphi_l| ))=1-\epsilon_l$  where $\lim_{l\to\infty}\epsilon_l=0$ and $\psi_{l}$ denotes the standard maximally entangled state on $\fr_l\otimes\fr_l$ with Schmidt rank $k_l$.
\end{enumerate}
Set $\rho^l:=\textrm{tr}_{\fr_l}( |\varphi_l\rangle\langle \varphi_l| )$ and
\[ \sigma_{\cn}^l:= id_{\fr_l}\otimes \crr_{\cn}^{l}\circ \cn^{\otimes l}(|\varphi_l\rangle\langle \varphi_l| ). \]
Then the data processing inequality and Lemma \ref{lemma:converse-uninformed-devetak} imply for each $\cn\in\fri$
\begin{eqnarray*}
  I_c(\rho^l,\cn^{\otimes l})&\ge& I_c(\rho^l, \crr_{\cn}^l\circ \cn^{\otimes l})\\
&=& \Delta (\sigma_{\cn}^l)\\
&\ge& \Delta (|\psi_{l}\rangle\langle \psi_{l}| )-\frac{2}{e}-8\log(k_l)\sqrt{\epsilon_l}\\
&=& \log k_l-\frac{2}{e}-8\log(k_l)\sqrt{\epsilon_l}.
\end{eqnarray*}
Consequently,
\begin{equation}\label{eq:ent-gen-6}
(1-8\sqrt{\epsilon_l})\frac{1}{l}\log k_l \le \frac{1}{l} \max_{\rho\in\mathcal{S}(\hr^{\otimes l})}\inf_{\cn\in\fri}I_c(\rho,\cn^{\otimes l})+ \frac{2}{le}
\end{equation}
and we end up with
\[R\le \limsup_{l\to\infty}\frac{1}{l}\log k_l \le \lim_{l\to\infty}\frac{1}{l} \max_{\rho\in\mathcal{S}(\hr^{\otimes l})}\inf_{\cn\in\fri}I_c(\rho,\cn^{\otimes l}), \]
which implies (\ref{eq:ent-gen-5}). The expression for $E_{IE}(\fri)$ is obtained in a similar fashion. We summarize the results in the following theorem.
\begin{theorem}[Entanglement-Generating Capacities of $\fri$] For arbitrary compound channels $\fri\subset\mathcal{C}(\hr,\kr)$ we have
\[E(\fri)=E_{ID}(\fri)=Q(\fri)=  \lim_{l\to\infty}\frac{1}{l} \max_{\rho\in\mathcal{S}(\hr^{\otimes l})}\inf_{\cn\in\fri}I_c(\rho,\cn^{\otimes l}), \]
and
\[E_{IE}(\fri)=Q_{IE}(\fri)= \lim_{l\to\infty}\frac{1}{l}\inf_{\cn\in\fri} \max_{\rho\in\mathcal{S}(\hr^{\otimes l})}I_c(\rho,\cn^{\otimes l}).  \]

\end{theorem}
\section{Conclusion and Further Remarks}

We have demonstrated that universal codes in the sense of compound quantum channels exist, and we determined the best achievable rates. The results are analogous to those well known related results from the classical information theory obtained by Wolfowitz \cite{wolfowitz-compound}, \cite{wolfowitz-book}, and Blackwell, Breiman and Thomasian \cite{blackwell}. In contrast to the classical results on compound channels there is, in general, no single-letter description of the quantum capacities for entanglement transmission and generation over compound quantum channels. Notice, however, that for compound channels with classical input and quantum output (cq-channels) a single-letter characterization of the capacity is always possible according to the results of \cite{bb-compound}.\\ 
Natural candidates of compound quantum channels that might admit a single-letter capacity formula are given by sets of quantum channels consisting entirely of degradable channels. While it is quite easy to see from the results in \cite{cubitt-ruskai-smith} that the degradable compound quantum channels with \emph{informed encoder} have a single-letter capacity formula for entanglement transmission and generation, the corresponding statement in uninformed case seems to be less obvious. This and related questions will be addressed in a future work.\\
Another issue we left open in this paper is the relation of the capacities considered here to other quantum communication tasks, for example to the subspace transmission and average subspace transmission and even to the randomized versions thereof. Again, we hope to come back to this point at some later time.

\appendix
\section{\label{sec:appendix}Appendix}
Let $\E$ and $\gr$ be subspaces of $\hr$ with
$\E\subset\gr\subset\hr$ where $k:=\dim \E$, $d_{\gr}:=\dim
\gr$. $p$ and $p_{\gr}$ will denote the orthogonal projections
onto $\E$ and $\gr$. For a Haar distributed random variable $U$
with values in $\mathfrak{U}(\gr)$ and $x,y\in\mathcal{B}(\hr)$ we
define a random sesquilinear form
\[b_{UpU^{\ast}}(x,y):= \textrm{tr}(UpU^{\ast} x^{\ast}UpU^{\ast}y )-\frac{1}{k}\textrm{tr}(UpU^{\ast}x^{\ast})\textrm{tr}(UpU^{\ast}y).  \]
In this appendix we will give an elementary derivation of the
formula
\begin{eqnarray}\label{eq:appendix-1}
  \mathbb{E}\{b_{UpU^{\ast}}(x,y) \}&=&  \frac{k^2-1}{d_{\gr}^2-1}\textrm{tr}(p_{\gr}x^{\ast}p_{\gr}y)+\frac{1-k^2}{d_{\gr}(d_{\gr}^2-1 )}\textrm{tr}(p_{\gr}x^{\ast})\textrm{tr}(p_{\gr}y)
\end{eqnarray}
for all $x,y\in \mathcal{B}(\hr)$ and where the expectation is taken with respect to the random variable $U$.\\
Let us set
\[ p_U:=UpU^{\ast}. \]
Since $ \textrm{tr}(p_Ux^{\ast}p_Uy) $ and $ \textrm{tr}(p_U
x^{\ast})\textrm{tr}(p_U y)$ depend sesquilinearly on
$(x,y)\in\mathcal{B}(\hr)\times \mathcal{B}(\hr)$ it suffices to
consider operators of the form
\begin{equation}\label{eq:appendix-2}
x=|f_1\rangle\langle g_1|\quad  \textrm{and}\quad
y=|f_2\rangle\langle g_2|
\end{equation}
with suitable $f_1,f_2,g_1,g_2\in\hr$. With $x,y$ as in
(\ref{eq:appendix-2}) we obtain
\begin{eqnarray}\label{eq:appendix-3}
  \textrm{tr}(p_Ux^{\ast}p_Uy)&=& \langle f_1,p_U f_2\rangle\langle g_2,p_U g_1\rangle\nonumber\\
&=& \langle f_1\otimes g_2, (U\otimes U)(p\otimes
p)(U^{\ast}\otimes U^{\ast})f_2\otimes g_1\rangle,
\end{eqnarray}
and
\begin{eqnarray}\label{eq:appendix-4}
 \textrm{tr}(p_U x^{\ast})\textrm{tr}(p_U y)&=& \textrm{tr}((p_U\otimes p_U)(|g_1\rangle\langle f_1|\otimes|f_2\rangle\langle g_2)   ) \nonumber\\
&=& \langle f_1\otimes g_2, (U\otimes U)(p\otimes
p)(U^{\ast}\otimes U^{\ast})g_1\otimes f_2\rangle.
\end{eqnarray}
Since the range of the random projection $(U\otimes U)(p\otimes
p)(U^{\ast}\otimes U^{\ast}) $ is contained in $\gr\otimes \gr$ we
see from (\ref{eq:appendix-3}) and (\ref{eq:appendix-4}) that we
may (and will) w.l.o.g. assume that $f_1,f_2,g_1,g_2\in \gr$.
Moreover, (\ref{eq:appendix-3}) and (\ref{eq:appendix-4}) show,
due to the linearity of expectation, that the whole task of
computing the average in (\ref{eq:appendix-1}) is boiled down to
the determination of
\begin{eqnarray}\label{eq:appendix-5}
 A(p)&:=& \mathbb{E}((U\otimes U)(p\otimes p)(U^{\ast}\otimes U^{\ast})   )\nonumber\\
&=& \int_{\mathfrak{U}(\gr)} (u\otimes u)(p\otimes
p)(u^{\ast}\otimes u^{\ast})du.
\end{eqnarray}
Obviously, $A(p)$ is $u\otimes u$-invariant, i.e. $A(p)(u\otimes
u)=(u\otimes u)A(p) $ for all $u\in\mathfrak{U}(\gr)$. It is
fairly standard (and proven by elementary means in \cite{werner})
that then
\begin{equation}\label{eq:appendix-6}
 A(p)=\alpha \Pi_s+\beta \Pi_a,
\end{equation}
where $\Pi_s $ and $\Pi_a$ denote the projections onto the
symmetric and antisymmetric subspaces of $\gr\otimes \gr$. More
specifically
\[\Pi_{s}:=\frac{1}{2}(\textrm{id}+\mathbb{F}) \qquad \Pi_a=\frac{1}{2}(\textrm{id}-\mathbb{F} ), \]
with $\textrm{id}(f\otimes g)=f\otimes g$ and $\mathbb{F}(f\otimes g)=g\otimes f$, for all $f,g\in\gr$.\\
Since $\Pi_s$ and $\Pi_a$ are obviously $u\otimes u$-invariant,
and $\Pi_s\Pi_a=\Pi_a\Pi_s=0$ holds, the coefficients $\alpha$ and
$\beta$ in (\ref{eq:appendix-6}) are given by
\begin{equation}\label{eq:appendix-7}
  \alpha=\frac{1}{\textrm{tr}(\Pi_s)}\textrm{tr}((p\otimes p)\Pi_s) =\frac{2}{d_{\gr}(d_{\gr}+1)}\textrm{tr}((p\otimes p)\Pi_s ),
\end{equation}
and
\begin{equation}\label{eq:appendix-8}
  \beta=\frac{1}{\textrm{tr}(\Pi_a)}\textrm{tr}((p\otimes p)\Pi_a) =\frac{2}{d_{\gr}(d_{\gr}-1)}\textrm{tr}((p\otimes p)\Pi_a ),
\end{equation}
where $d_{\gr}=\dim \gr$ and we have used the facts that
\[ \textrm{tr}(\Pi_s)=\dim \textrm{ran}(\Pi_s)=\frac{d_{\gr}(d_{\gr}+1)}{2}  \]
and
\[  \textrm{tr}(\Pi_a)=\dim \textrm{ran}(\Pi_a)=\frac{d_{\gr}(d_{\gr}-1)}{2}. \]
It is easily seen by an explicit computation with a suitable basis
that
\begin{equation}\label{eq:appendix-8a}
 \textrm{tr}((p\otimes p)\Pi_s )=\frac{1}{2}(k^2+k)\quad \textrm{and}\quad \textrm{tr}((p\otimes p)\Pi_a )=\frac{1}{2}(k^2-k).
\end{equation}
For example choosing any orthonormal basis $\{e_1,\ldots,
e_{d_{\gr}}  \}$ of $\gr$ with $e_1,\ldots,e_k\in\textrm{ran}(p)$
we obtain
\begin{eqnarray*}
  \textrm{tr}((p\otimes p)\Pi_s )&=& \sum_{i,j=1}^{d_{\gr}}\langle e_i\otimes e_j, (p\otimes p)\Pi_s e_i\otimes e_j\rangle\\
&=& \sum_{i,j=1}^{k}\langle e_i\otimes e_j, (p\otimes p)\Pi_s e_i\otimes e_j\rangle\\
&=& \frac{1}{2}\Big(\sum_{i,j=1}^{k}\langle e_i,e_i\rangle\langle e_j,e_j\rangle+\langle e_i,e_j\rangle\langle e_j,e_i\rangle              \Big)\\
&=& \frac{1}{2}(k^2+k),
\end{eqnarray*}
with a similar calculation for $\textrm{tr}((p\otimes p)\Pi_a ) $.
Utilizing (\ref{eq:appendix-7}), (\ref{eq:appendix-8}),
(\ref{eq:appendix-8a}), and (\ref{eq:appendix-6}) we end up with
\begin{equation}\label{eq:appendix-9}
  A(p)=\frac{k^2+k}{d_{\gr}(d_{\gr}+1)}\Pi_s+ \frac{k^2-k}{d_{\gr}(d_{\gr}-1)}\Pi_a.
\end{equation}
Now, (\ref{eq:appendix-9}), (\ref{eq:appendix-5}),
(\ref{eq:appendix-4}), (\ref{eq:appendix-3}), and some simple
algebra show that
\begin{eqnarray*}
\mathbb{E}\{ \textrm{tr}(UpU^{\ast} x^{\ast}UpU^{\ast}y )-\frac{1}{k}\textrm{tr}(UpU^{\ast}x^{\ast})\textrm{tr}(UpU^{\ast}y)\}&=&\frac{k^2-1}{d_{\gr}^2-1}\textrm{tr}(x^{\ast}y)\\
& &+\frac{1-k^2}{d_{\gr}(d_{\gr}^2-1
)}\textrm{tr}(x^{\ast})\textrm{tr}(y).
\end{eqnarray*}
\emph{Acknowledgment.} 
We would like to thank Mary Beth Ruskai and the referee for many helpful suggestions and advices that led to significant improvement of the overall structure and readability of the paper.\\
I.B. is supported by the Deutsche
Forschungsgemeinschaft (DFG) via project ``Entropie und
Kodierung gro\ss er Quanten-Informationssysteme'' at the TU
Berlin. H.B. and J.N. are grateful for the support by TU
Berlin through the fund for basic research.


\end{document}